\documentclass{aa}  

\usepackage{graphicx}
\usepackage{txfonts}
\usepackage{lipsum}
\usepackage{subcaption}         % necessary for continued figures, example in section 3
\usepackage{lscape}             % to rotate a single page table, example in appendix.
\usepackage{placeins}           % useful with \FloatBarrier, to keep 
\usepackage{amsmath}
\usepackage{verbatim}
\usepackage{xcolor}
\usepackage{footnote} 
\usepackage{multicol}
\usepackage{multirow}
\usepackage{tabularx}
\usepackage{soul}
\usepackage{natbib}
\bibpunct{(}{)}{;}{a}{}{,} 

\usepackage[colorlinks=true,backref=page]{hyperref}
\hypersetup{colorlinks,breaklinks,citecolor=blue, urlcolor=blue,linkcolor=blue}
\newcommand{\HR}{HR\,4796}

\begin{document}

   \title{The dust in Sauron's eye\thanks{\href{https://www.newscientist.com/article/dn25676-eye-of-sauron-star-spotted-by-planet-hunting-camera/}{{https://www.newscientist.com/article/dn25676-eye-of-sauron-star-spotted-by-planet-hunting-camera/}}}}%\footnote{https://www.newscientist.com/article/dn25676-eye-of-sauron-star-spotted-by-planet-hunting-camera/}%
   \subtitle{Observational and experimental results on the debris disk around \HR}
%%%%%%%%%%%%%%%%%%%%%%%%%%%%%%%%%%%%%%%%
% Please do not include ORCIDs next to author names.
% Only ORCIDs authenticated by individual authors in EDP Sciences editorial system will be taken into account.
% ORCIDs included here will be removed.
%%%%%%%%%%%%%%%%%%%%%%%%%%%%%%%%%%%%%%%%

   \author{M. Bonduelle\inst{1}
        \and J. Milli\inst{1} \and O. Poch\inst{1} \and N. Engler\inst{2} \and  M. Rigouleau\inst{1} \and R. Tazaki\inst{3} \and J. Ma\inst{1} \and J. Olofsson\inst{4}  \and J-B. Renard\inst{5} \and J. Lasue\inst{6} \and R. Sultana\inst{7} \and G. Duchêne\inst{1,8} \and L. Martinien\inst{1} \and M. Roumesy\inst{1} \and C.H. Chen \inst{9,10} \and C. Lisse\inst{11} \and J-C. Augereau\inst{1}%\thanks{Shows the usage of elements in the author field}
        }

   \institute{Univ. Grenoble Alpes, CNRS, IPAG, 38000 Grenoble, France\\
            \email{myriam.bonduelle@univ-grenoble-alpes.fr}
            \and ETH Zurich, Institute for Particle Physics and Astrophysics, 3688 Wolfgang-Pauli-Strasse 27, CH-8093 Zurich, Switzerland
            \and Department of Earth Science and Astronomy, The University of Tokyo, Tokyo 153-8902, Japan
            \and European Southern Observatory, Karl-Schwarzschild-Strasse 2, 85748 Garching bei München, Germany
            \and LPC2E-CNRS, Orléans, France
            \and IRAP, Université de Toulouse, CNES, CNRS, UPS, Toulouse, France
            \and LIRA, Observatoire de Paris/PSL, Sorbonne Université, Université Paris Cité, CY Cergy Paris Université, CNRS, 5 place Jules Janssen, 92190 Meudon, France  
            \and Astronomy Department, University of California Berkeley, Berkeley CA 94720-3411, USA
            \and William H. Miller III Department of Physics and Astronomy, John's Hopkins University, 3400 N. Charles Street, Baltimore, MD 21218, USA
            \and Space Telescope Science Institute, 3700 San Martin Drive, Baltimore, MD 21218, USA
            \and Space Department, Johns Hopkins University Applied Physics Laboratory, 11100 Johns Hopkins Rd, Laurel, MD, USA
}

   \date{Received September 30, 20XX}

% \abstract{}{} {}{}{}
% 5 {} token are mandatory
 
  \abstract
   {The debris disk surrounding \HR{} is a bright, narrow ring observed at multiple wavelengths, both in scattered light and thermal emission. The optical properties of the dust particles (scattering phase function, SPF; degree of linear polarization, DoLP; reflectance) that can be retrieved through scattered light imaging are linked to the physicochemical properties of the dust particles (size, shape, composition, etc.), but the retrieving process is challenging. Among debris disks, \HR{} presents several peculiarities, in particular unusually high DoLP values at small scattering angles.}
   {This study is aimed at improving the constraints on the physicochemical properties of the dust particles orbiting in the disk.}  
   {We used multiwavelengths scattered light observations of \HR, obtained with SPHERE/IRDIS and SPHERE/ZIMPOL. We forward modeled those observations using a novel joint parametric approach to constrain the morphology, the scattering phase function and the degree of linear polarization of the disk. These observations are then compared to laboratory measurements, previous observations of \HR{} and of some Solar System asteroids and comets.}
   {We present new observational data in scattered light of \HR, for $\lambda$ = 0.63 µm, $\lambda$ = 0.79 µm, and $\lambda$ = 1.25 µm. We modeled the disk at each of these wavelengths and found geometric parameters consistent with previous results. We obtained the parametrized SPF and DoLP over the whole range of scattering angles (13° to 167°) and extracted the DoLP at 90° (without any parametrization) as well as the spectral reflectance from the visible to near-infrared (NIR) wavelength range. We compared the reflectance and polarimetric properties of HR 4796 to those measured on a laboratory dust sample and obtained from previous results on \HR. We find that the sample providing the best match to these properties appears to be large (from a few micrometers to 100 µm) iron sulfides particles, such as pyrrhotite and troilite. Such iron sulfides are among the opaque minerals (Fe\textsubscript{x}S\textsubscript{y}, FeNi, etc.) that contribute to the low albedo of some small bodies of the Solar System, such as B-,C-,D, P-type asteroids, comets, and so on.}
   {We confirm that the DoLP of \HR{} peaks at high values (> 45\%) and for small scattering angles (< 55°). At a 90° scattering angle, we observe a red spectral slope in the visible and NIR wavelength range, as well as a blue polarimetric slope. These results are compatible with large (a few micrometers to 100 µm) absorbents being the main scatterers in the disk. We show that the DoLP of \HR{} is notably different from that of Solar System comets and seems closer to that of some more processed near-Earth orbit asteroids, possibly indicating devolatilization and/or melting induced by space weathering on the dust particles of \HR.}


   \maketitle
\nolinenumbers

%%%%%%%%%%%%%%%%%%%%%%%%%%%%%%%%%%%%%%%%%%%%%%%%%%%%%%%%%%%%%%
\section{Introduction}
The formation and evolution of planetary systems is traced by circumstellar disks, which are rings of material orbiting a central star. In the first stages of planetary formation, it takes the form of a massive dust- and gas- rich belt called a protoplanetary disk. As it evolves, the gas is depleted and the small dust particles grow into planetesimals (km-sized rocky and icy bodies) and, occasionally, planets. The outcome of this evolution are debris disks, analogous to the main asteroid belt and Kuiper belt, and final stages of circumstellar disks. These disks are dominated by destructive processes, and continuously replenished in small dust by the collisions of the bodies they contain, which grind the material into micro- or milli-metric particles. Debris disks are therefore considered to be the markers of successful planetesimal and possibly planetary formation \citep{Krivov_2010}, and the dust they contain is an indicator of the composition of planetesimals or planets material \citep{Hughes_2018}.\\
The thermal emission of the millimetric dust contained in debris disks can be seen using high angular resolution observations in the submillimeter (submm) range, whereas the scattering properties of the micron-sized particles can be observed through high contrast imaging in the visible and near-infrared (NIR). In this regime, the Spectro-Polarimetric High-contrast Exoplanet REsearch (SPHERE) instrument \citep{Beuzit_2019}, installed at the Very Large Telescope (VLT), has made it possible to carry out recent works aimed at imaging and resolving numerous debris disks \citep{Engler_2025}. These efforts have helped in characterizing the morphology and optical properties of those disks. \\
\HR{} is an early-type star (A0V), with an age estimated at 10 $\pm$ 3 Myr-old \citep{Bell_2015}, located at $\sim$ 70.8 pc \citep{Gaia_2023}, orbited by a bright, narrow, and highly inclined ($\sim$76\textdegree) debris disk \citep{Milli_2017}. This debris disk has been extensively studied, both in the submm regime \citep{Telesco_2000, Kennedy_2018} and in scattered light \citep{Perrin_2015,Milli_2017,Chen_2020}; it also shows an extended halo of submicron-sized particles \citep{Schneider_2018}. The disk presents several interesting peculiarities, starting with its narrowness, which is unexpected in a young debris disk \citep{Lisse_2017,Kueny_2026}, a two-component scattering phase function that is unusual among debris disks \citep{Engler_2023} and that cannot be reproduced by simple Mie or distribution of hollow spheres (DHS) models \citep{Milli_2017}, and a high degree of linear polarization at low scattering angles \citep{Arriaga_2020}. Despite all of these observations, the composition of the dust in the disk remains an open question. Dust particles made of amorphous silicates and carbons, as well as water ice (common species in interstellar dust) were proposed by \citet{Augereau_1999, Li_2003, Kohler_2008}; whereas \citet{Debes_2008} suggested a dust composition dominated by 1.4\,µm grains made of complex organic matter similar to laboratory tholins produced by irradiation of simple molecules in gas or ice mixtures. \citet{Lisse_2017} proposed a composition resembling certain objects of the Solar System, namely, made up of processed, devolatilized, large, porous aggregates. \\
As the scattering properties of a material depend on its physical properties (composition, structure, size, etc.), debris disks observations in scattered light allow us to constrain the nature of the dust particles and trace back the composition of the planetesimals and planets that formed them. However, this is a degenerate problem and multiple sources of information (multiwavelengths, polarized light observations, etc.) are required to break some of these degeneracy. In that context, laboratory measurements on samples of a controlled size and composition offer a possible avenue of exploration for different particle compositions in debris disks. 
In this paper, we bring new insights on the disk's composition, by combining several sources of information and to hopefully break some of the degeneracy of the problem. We compare the optical and NIR properties of the debris disks orbiting around \HR, both in terms of the total intensity and polarized light. From these multiwavelength observations, we extracted the scattering phase function, degree of linear polarization, and spectral reflectance as a function of the wavelength. We then compared these results to previous observations of \HR{}, as well as to laboratory measurements made on an iron sulfide sample and to observations of Solar System objects. \\
We present the observations and the data reduction in Sect.\,\ref{sec_obs}, followed by the foward modeling strategy in Sect.\,\ref{sec_fm} and the results we obtained in Sect.\,\ref{results_models}, both from the models (\ref{results_morpho} and \ref{results_phf}) and from the aperture photometry procedure we carried out on the data directly (\ref{subsec_90}). We present our laboratory measurements in Sect.\,\ref{sec_lab_mes} and a discussion and summary of our results in Sect.\,\ref{sec_discussion}.

%%%%%%%%%%%%%%%%%%%%%%%%%%%%%%%%%%%%%%%%%%%%%%%%%%%%%%%%%%%%%%
\section{Observations}
\label{sec_obs}
\subsection{Instrumental setup}
\label{subsec_setup}
\HR{} was observed with the SPHERE instrument in linear polarization, both in the optical with the ZIMPOL subsystem using the broadband filters R\_PRIM and I\_PRIM (central wavelengths $626.3$ nm and $789.7$ nm), and in the NIR with the IRDIS subsystem using the broadband J filter centered at $1245$ nm. The log of the observations is given in Table \ref{tab_log}, summarizing the setup and the atmospheric and turbulence conditions. 

With ZIMPOL, the observations used the slow pol detector readout mode with the minimum detector integration time (DIT) of 10 s, allowing for a higher sensitivity thanks to a lower readout noise. The observations did not use any coronagraph and the central part of the point spread function (hereafter, PSF) was purposely heavily saturated. This strategy was already tested on previous ZIMPOL observations of this target \citep{Milli_2019} and we chose to reapply it here because the ZIMPOL detector is not very sensitive to saturation and it allows to better correct the beamshift effect \cite[see][for a description of this effect]{Schmid2018}. We obtained 40 deep polarization cycles that were interleaved at the beginning, middle, and end by a short unsaturated fast pol polarization cycle (with a DIT of 2\,s and the neutral density filter ND\_1). We used these cycles to measure the photometry of the star and retrieve the accurate position of the star center. The observations were carried out in the p1 mode, where all optics inside the instrument were fixed (e.g., the derotator) to better calibrate the instrumental polarization. In this setup, both the field of view and the pupil were rotating. 

With IRDIS, we observed the star in pupil tracking mode, with the N\_ALC\_YJH\_S coronagraph with a diameter of 92 mas and a DIT of 32\,s. The coronagraphic observations consisted in eight polarization cycles. A waffle pattern applied on the deformable mirror before and after the deep coronagraphic sequence created four PSF echoes that are used to measure the center of the star behind the coronagraphic mask. A noncoronagraphic image of the star was also obtained with a DIT of 6\,s and the neutral density filter ND\_2, which  allowed us to measure the PSF and retrieve the photometry of the star. The last sky background observations were obtained at the end of the observations.

\begin{table*}
\caption{Log of the SPHERE observations of \HR\,A.}
\label{tab_log}
\centering
\begin{tabular}{l c c c c c c c c}
\hline 
Night & Subsystem & Filter & DIT x NDIT x &    Paral.  & Seeing &        Coh. &       Wind\tablefootmark{d}   & Strehl\tablefootmark{e}  \\
 & &  &  NHWP x NCYC\tablefootmark{a} &   angle\tablefootmark{b} ($^\circ$)  &  (")  &  time\tablefootmark{c} (ms) & (m/s)  & (\%)  \\
\hline
\hline
2020-02-23 & ZIMPOL & R\_PRIM & 10x2x4x40 & $69.2$ & $0.5\pm0.1$ & $10\pm2$ & 4-7 &  24 \\
2020-03-19 & ZIMPOL & I\_PRIM & 10x2x4x40 & $73.7$ & $0.8\pm0.2$ & $5\pm1$ & 7-10 & 25 \\
2021-04-07 & IRDIS & BB\_J & 32x4x4x8  & $58$ & 0.5 & 8.4 &  & $76\pm3$ \\
\hline
\end{tabular}
\tablefoot{
\tablefoottext{a}{DIT is the individual detector integration time in second. NDIT is the number of DIT for a given position of the Half-Wave-Plate (HWP). NHWP is the number of positions of the HWP for a given polarimetric cycle (4 in our case for $0^\circ$,$90^\circ$,$45^\circ$, and $135^\circ$ to retrieve both Stokes Q and U). NCYC is the number of HWP cycles.}
\tablefoottext{b}{Parallactic angle variations during the observations}
\tablefoottext{c}{Coherence time $\tau_0$ as measured by the MASS-DIMM in milliseconds.}
\tablefoottext{d}{The wind speed as measured by the Astronomical Site Monitor (ASM) meteo tower at a height of 30m.} 
\tablefoottext{e}{The Strehl as estimated by the real-time computer SPARTA and converted to the central wavelength of each filter.}
}
\end{table*}

\subsection{Data reduction}

\begin{figure*}[ht!]
\centering
\includegraphics[width=0.96\textwidth]{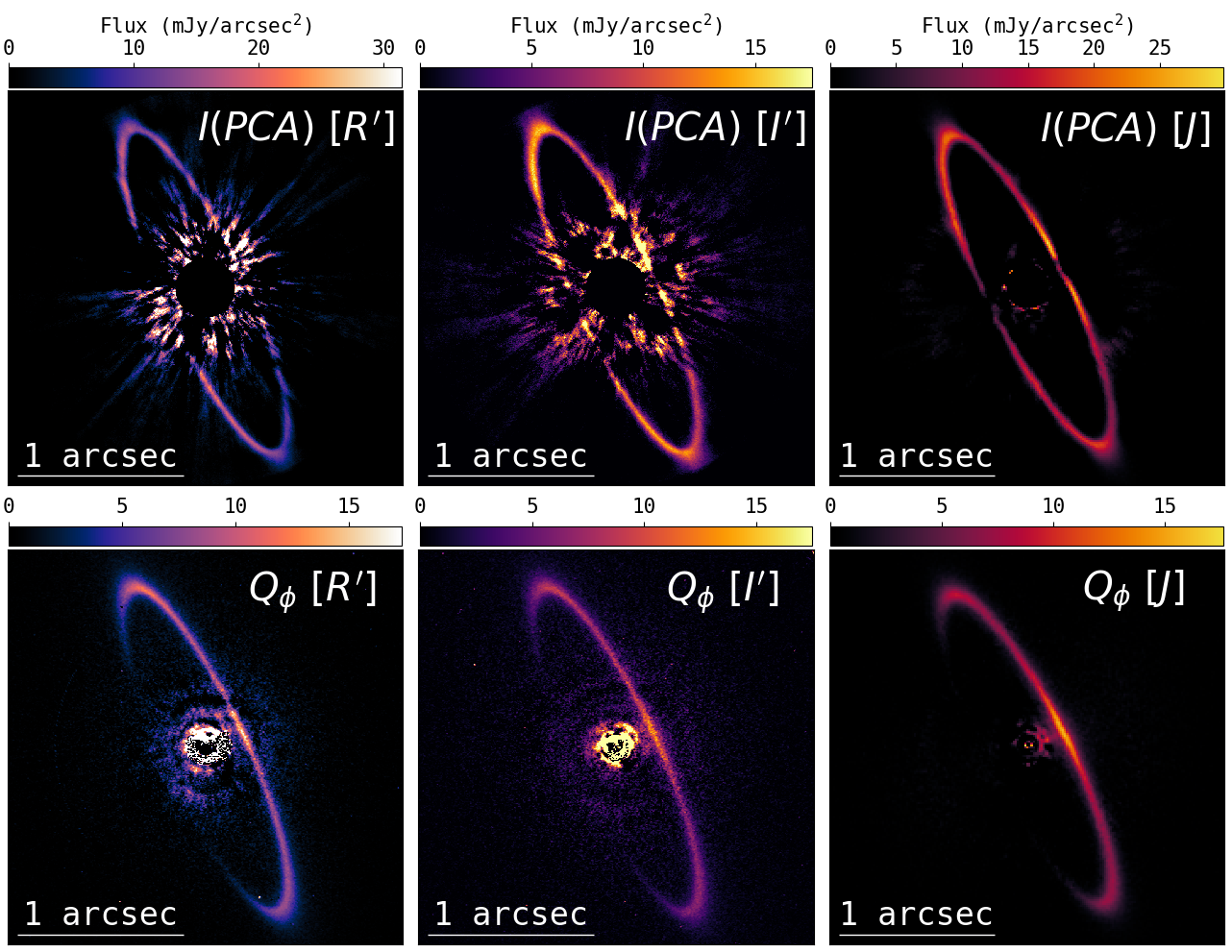}
    \caption{Post-processed observational datasets in mJy/arcsec$^2$. Top row: Total intensity images (principal component analysis, with $n_{comp} = 4$ for ZIMPOL and $n_{comp} = 2$ for IRDIS). Bottom row: Polarized intensity images. From left to right: ZIMPOL R' band (centred around 0.63 µm), ZIMPOL I' band (0.79 µm), IRDIS J band (1.25 µm).
    }
    \label{fig_Obs_all}
\end{figure*}

The ZIMPOL observations were processed by the High-Contrast Data Centre\footnote{\href{https://hc-dc.cnrs.fr}{\url{https://hc-dc.cnrs.fr}}} \cite[HC-DC,][]{Delorme2017}, which implements the SPHERE-ZIMPOL (SZ) software package developed at ETH Zurich and described in \citet{Schmid2017}. It extracts the Stokes, I, Q, and U images from the sequence of polarimetric cycles. Stokes Q and U can be derived using polarimetric differential imaging (PDI). To obtain the best-quality polarized intensity image, an accurate recentering of the star is needed between the orthogonal linear polarization channels, encoded in the even and odd rows of the ZIMPOL detector \citep{Schmid2018}. This subpixel recentering is required due to the beamshift effect, described in \citet{vanHolstein_2023}. Following \citet{Milli_2019}, we found that using the barycenter of the saturated pixels to recenter the images provided the best stellar attenuation and correction of the beamshift. The Stokes Q and U images were further converted by the HC-DC workflow into the azimuthal Stokes parameters $Q_\phi$ and $U_\phi$,  defined as $Q_\phi = -Q\cos(2\phi) - U\sin(2\phi)$ and $U_\phi = +Q\sin(2\phi) - U\cos(2\phi)$  \citep{deBoer_2020}, where $\phi$ is the polar angle between north and the point of interest measured from the north over east (the position angle). The polarized signal of the disk is contained in the $Q_\phi$ images visible in the bottom row of Fig.\,\ref{fig_Obs_all}\footnote{The ZIMPOL R' and IRDIS colormaps used throughout the paper come from the cmasher package \citep{vanderVelden_2020}}  (left and middle). It is positive, indicating a tangential polarization. A very faint signal is detectable in the $U_\phi$ images, which we attribute to the effect of the convolution of the astrophysical signal with the PSF, occurring even if $U_\phi$ is zero before convolution, as shown in Appendix A of \citet{Engler2018}.\\
The pipeline also delivers a temporal sequence of Stokes I images, corresponding to the total intensity images of the star. In the ZIMPOL p1 mode, both the field of view and the pupil are rotating. This means that both the disk and the spider diffraction pattern of the telescope are rotating. However, the most prominent PSF features are temporally stable in this configuration: the speckle ring near the adaptive optics control radius, the strong fixed speckles due to the imprint of the deformable mirror actuator grid and the vertical frame transfer trail above and below the saturated pixels. Therefore, we treated this temporal cube in a way similar to a cube obtained in pupil-tracking mode, applying angular differential imaging techniques \cite[ADI,][]{Marois2006} to subtract the PSF and reveal the circumstellar environment. We used the principal component analysis implementation of ADI \citep{Soummer2012,Amara2012}, subtracting four principal components. The PCA-reduced Stokes I ZIMPOL images of the disk are visible in the top row of Fig.\ref{fig_Obs_all} (left and middle). The disk is resolved in our observations (as previously observed in \citealp{Milli_2017}), and the innermost disk regions within 0.5\arcsec{} are not clearly detected because of the strong speckles still present and the disk self-subtraction in ADI \citep{Milli2012, Juillard_2023}. \\
For IRDIS, we used the IRDAP pipeline \citep{vanHolstein_2020_irdap,vanHolstein_2020} to obtain the Stokes images $I$, $Q_\phi$, and $U_\phi$ and to correct for the instrumental polarization effects of the complete optical system. The polarization of the central stellar halo was estimated to $0.08\%\pm 0.1\%$ at the $1\sigma$ level by the pipeline. Therefore the star can be considered unpolarized, but we prefer using the image subtracted from this tiny stellar polarization to have a slightly enhanced data quality. The version of IRDAP was slightly modified to avoid averaging the NDIT=4 images acquired with the same position of the HWP. This yielded a temporal sequence of 128 Stokes I images, which were post-processed with a PCA algorithm to remove the stellar halo as done with ZIMPOL. We empirically determined that removing two principal components yielded the best signal-to-noise ratio (S/N) on the disk. The IRDIS Stokes I and $Q_\phi$ images are visible in the rightmost column of Fig.\ref{fig_Obs_all}.\\
All the data presented on Fig.\ref{fig_Obs_all} are shown in mJy/arcsec$^2$. The images are normalized to the stellar flux, with its value in Jy extracted from the SIMBAD database ([R'] = 15.8 Jy; [I'] = 12.5 Jy; and [J] = 7.67 Jy) and expressed by pixel surface. The S/N maps $[data/noise]$ obtained for these data are presented in the top rows of Fig.\ref{fig_SNR_tot} and Fig.\ref{fig_SNR_pol} in Appendix \ref{app_res_best_models}, showing that the IRDIS data yield the best S/N ($\sim$15 in the ansae, against $\sim$10 for both Zimpol R' and I').

%%%%%%%%%%%%%%%%%%%%%%%%%%%%%%%%%%%%%%%%%%%%%%%%%%%%%%%%%%%%%%
\section{Forward modeling procedure} 
\label{sec_fm}
\subsection{Model description}
\label{subsec_mod_descr}
To constrain the disk properties simultaneously for the total intensity and polarized intensity data, we carried out a foward modeling of the observations using a simple disk model. 
A given set of parameters (presented hereunder) were injected in the scattered light disk module of the \href{https://vip.readthedocs.io/en/latest/vip_hci.fm.html}{\texttt{VIP/HCI}} forward modeling package \citep{Gomez_Gonzalez_2017,Christiaens_2023} to generate a synthetic disk. This synthetic disk was then tested against the data, using a reduced chi squared $\chi^2_r$ as a metric to assess the validity of this parameter set. The process was iterated so as to converge towards the best model. For each of the wavelengths, the final models presented in \ref{results_models} are so-called joint models, meaning that during the foward modeling process, each parameter set was tested simultaneously on the total intensity and polarized intensity images, with common morphological parameters, as well as separated phase functions parameters.
The parameter space was first explored by using a Nelder-Mead algorithm to rapidly converge to a minimum (\href{https://docs.scipy.org/doc/scipy-1.17.0/reference/optimize.html}{\texttt{scipy.optimize}} package \citealp{Virtanen_2020}), allowing us to obtain a correct first guess. Then we applied a Markov chain Monte Carlo framework, using the \href{https://github.com/dfm/emcee?tab=readme-ov-file}{\texttt{emcee}} package \citep{Foreman_Mackey_2013}. 

\subsubsection{Morphology}
\label{model_morpho}
The morphological description of our models of the disk is detailed in Appendix \ref{app_disk_morpho}. We followed previous descriptions of the disk, assuming an elliptical disk \citep{Milli_2017, Milli_2019}, with $a$ as the semi-major axis of the disk, $e$ the ellipticity, $i$ the inclination, $PA$ the position angle, $\omega$ the argument of the pericenter, and $\Psi$ the disk vertical opening angle.\\
The dust volume density distribution was parametrized using a smoothly connected double power law \citep{Augereau_1999}, with $\alpha_{in}$ and $\alpha_{out}$ the inner and outer slopes (respectively). In all our models, the inner slope of the dust distribution, $\alpha_{in}$ is fixed at 30 (further discussed in Appendix \ref{app_alphain}) and in the ZIMPOL models, the opening angle is $\Psi = 1.2\%$; whereas it was left as a free parameter in the IRDIS models (see Appendix \ref{app_opang}). There are consequently seven (or six when $\Psi$ is fixed) morphological free parameters in our model.

\subsubsection{Phase function}
\label{model_phf}
To complete the disk's modeling, the scattering phase function, both in total intensity (SPF) and polarized intensity (pSPF), need to be determined. At a given wavelength, the (p)SPF represents the variation of the (polarized) intensity of the starlight scattered by the dust grains as a function of the scattering angle, with the accessible range of scattering angles depending on the geometry of the disk \citep{Perrin_2015}. The shape of the SPF and of the degree of linear polarization (DoLP), defined as $DoLP = \frac{pSPF}{SPF}$ strongly depend on the composition, shape, size, and porosity of the dust grains. To derive the (p)SPF from the morphology presented in \ref{model_morpho}, the dust properties must be azimuthally uniform, with no significant asymmetries observed in µm-sized dust in the disk \citep{Milli_2019}.\newline 
To describe the total intensity flux and the polarized intensity one, we used a parametrization of the SPF and of the DoLP (this parametric approach is discussed in Appendix \ref{param_descr_phf}). For the scattering phase function, we used a two-component Henyey-Greenstein \citep{Hong_1985,Hapke_2012, Milli_2017}, defined as
\begin{equation}  
\begin{split}
SPF \propto \Bigg( \Bigg. w\; \frac{1-g_1^2}{ \left[1+g_1^2 -2\cdot g_1\cdot cos(\theta)\right]^{3/2}}\; \\ +\; (1-w) \; \frac{1-g_2^2}{\left[1+g_2^2 -2\cdot g_2\cdot cos(\theta)\right]^{3/2}} \Bigg. \Bigg).
\end{split}
\label{eqDHG}
\end{equation}

with $g_1$, $g_2$ the coefficients for each of the Henyey-Greenstein components, $w$ the weight of each of the components and $\theta$ the scattering angle. \newline
For the degree of linear polarization, we used the parametrization proposed in \citet{Ren_2023} for protoplanetary disks, defined through the so-called beta function. This parametric description depends on three parameters, $\alpha$, $\beta$, and $f_M$, expressed as

\begin{equation}  
DoLP = f_M \; \frac{B\left(\frac{\theta}{\pi} \mid \alpha, \beta \right)}{B(\frac{\alpha - 1}{\alpha + \beta - 2}\mid \alpha, \beta)}. 
\label{eqDolp}
\end{equation}

\noindent The $B$ function is defined by:

\begin{equation}  
B(x\mid \alpha, \beta) = \frac{\Gamma(\alpha + \beta)}{\Gamma(\alpha) + \Gamma(\beta)} \cdot x^{\alpha - 1} (1-x)^{\beta -1}
\label{eqDolp1},
\end{equation}
with $\Gamma(x) = \int^{\infty}_{0} t^xe^{-t}$. \\

For the joint models, this amounts to 14 free parameters in total: seven (or six) morphological parameters (see \ref{model_morpho}); four parameters describing the scattering phase function ($g_1$, $g_2$, $w$, $sc$); and three describing the degree of linear polarization ($\alpha$, $\beta$, $f_M$).

\subsection{Markov chain Monte Carlo method}
\label{sec_mcmc}

\begin{figure*}[ht!]
\centering
\includegraphics[width=0.96\textwidth]{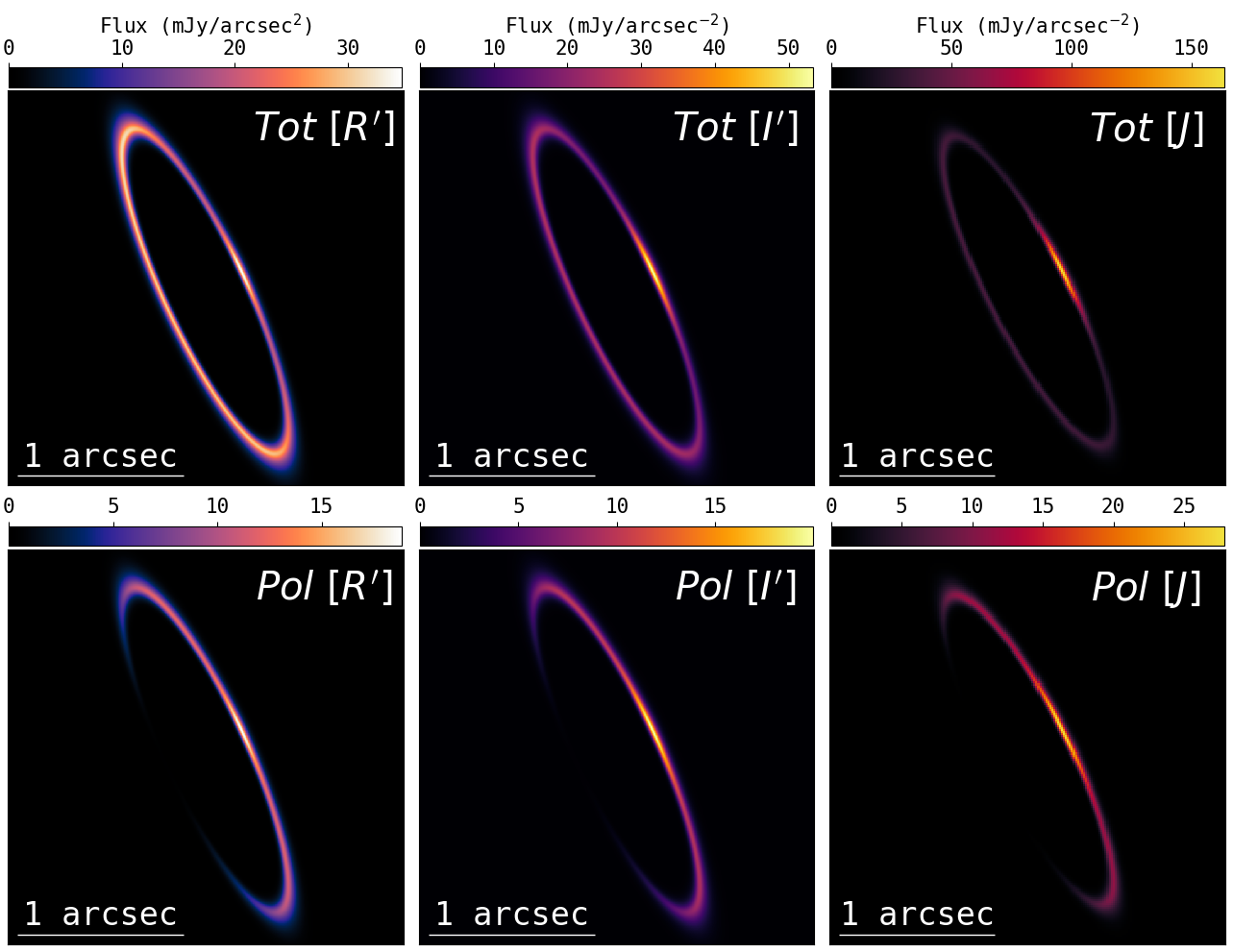}
    \caption{Best foward modeling resulting from the MCMC in mJy/arcsec$^2$, unconvolved. Top row: Total intensity images, bottom row: Polarized intensity images. From left to right: ZIMPOL R' band (0.63 µm), ZIMPOL I' band (0.79 µm), IRDIS J band (1.25 µm).}
    \label{fig_Best_Mod}
\end{figure*}

To refine our model and explore the parameter space, we used an MCMC approach to minimize the residuals between the observation and the generated model \cite[\href{https://github.com/dfm/emcee?tab=readme-ov-file}{\texttt{emcee}} package,][]{Foreman_Mackey_2013}. Following \citet{Hogg_2010}, the likelihood function is defined as $e^{-\chi^2/2}$, with the chi-squared $\chi^2$ defined as the squared ratio between the residuals and the noise map.\\ 
Using the results from section \ref{model_phf}, we kept a parametric description for both the scattering phase function and the degree of linear polarization, allowing us to reduce the number of parameters. The morphology of the disk is still described through the same seven (or six) parameters, with the SPF described using four parameters ($g1$, $g2$, $w$, and $sc$), and the DoLP using three ($\alpha$, $\beta$, $f_M$). As described previously, for each iteration of the MCMC, the parameter set is injected in the scattered light module of the \href{https://vip.readthedocs.io/en/latest/vip_hci.fm.html}{\texttt{VIP/HCI}} foward modeling package that we modified to generate the disk model through a parametric description of the SPF and DoLP rather than SPF and pSPF.
For each of the models, we used the first-guesses previously obtained with a Nelder-Mead algorithm (\ref{subsec_mod_descr}). Except for the opening angle $\Psi \geq 1.2\%$ in the IRDIS joint modeling, no constraints were placed on the values of morphological parameters of the model. The phase function parameters values were constrained using their intrinsic definitions: 
\begin{itemize}
\begin{multicols}{3}
    \item[] $ -1 \leq g_1;g_2 \leq 1$ 
    \item[] $~~~1\leq \alpha;\beta$
    \item[] $0\leq w \leq 1$
    \item[] $0 \leq f_{M} \leq 1$ 
\end{multicols}
\end{itemize}
Due to the size of the data cubes and the available resources, each MCMC iteration took on average $\sim$64 seconds for IRDIS; $\sim$260 seconds for ZIMPOL I' and $\sim$330 seconds for ZIMPOL R'. This resulted in adjustments regarding the number of parallel walkers (100 for IRDIS, 60 for ZIMPOL) and iterations (2 000 for IRDIS and ZIMPOL I' and 1 541 for ZIMPOL R'). After the MCMC converged, we removed some of the initial iterations (burn-in) and extracted the best model and uncertainties using the 50\textsuperscript{th} (median), 0.1\textsuperscript{th} and 99.9\textsuperscript{th} percentiles. \\
The best models obtained are shown in Fig. \ref{fig_Best_Mod}, with the top-row corresponding to the total intensity and the bottom row to the polarized light. The corresponding $\chi^2_r$ are shown in Table \ref{table_morph_param}, and the S/N map of the residuals $[(data-best ~model)/noise]$ are shown in Fig. \ref{fig_SNR_tot} and Fig.\,\ref{fig_SNR_pol} bottom rows in Appendix \ref{app_res_best_models}. \\
The morphological results obtained are presented in \ref{results_morpho} and the phase functions in \ref{results_phf}. 

\section{Results}
\label{results_models}

\subsection{Morphology of the disk}
\label{results_morpho}
The results presented in Table \ref{table_morph_param} correspond to the best-fit parameters (50th percentile) from our three models, after discarding the burn-in phase. The uncertainties are estimated from the 0.99\textsuperscript{th} and 0.01\textsuperscript{th} quartiles (at 3$\sigma$).

\begin{table}[ht!]
\renewcommand{\arraystretch}{2} % Default value: 1
\caption{Morphological parameters obtained for the ZIMPOL and IRDIS models.}                 % title of Table
\label{table_morph_param}   
\centering                        
\begin{tabular}{c | c | c | c}      
\hline               
 & ZIMPOL (R') & ZIMPOL (I') & IRDIS (J) \\         
\hline
$\lambda_c$ & 0.63 µm & 0.79 µm & 1.25 µm  \\         
\hline                      
$a$ (mas)      & $1064.5^{+0.4}_{-0.4}$    & $1066.0^{+0.5}_{-0.5}$    & $1063.4^{+0.5}_{-0.5}$    \\    
$PA$ (°)       & $26.68^{+0.01}_{-0.01}$   & $26.76^{+0.01}_{-0.01}$   & $26.74^{+0.02}_{-0.01}$   \\
$i$ (°)        & $76.61^{+0.01}_{-0.01}$   & $76.61^{+0.02}_{-0.02}$   & $76.70^{+0.02}_{-0.02}$   \\
$e$            & $0.045^{+0.001}_{-0.001}$ & $0.045^{+0.001}_{-0.001}$ & $0.064^{+0.001}_{-0.001}$ \\    
$\omega$ (°)       & $-110.96^{+0.39}_{-0.35}$ & $-112.55^{+0.51}_{-0.47}$ & $-108.15^{+0.37}_{-0.39}$  \\    
$\Psi (\%)$    & $1.20$ [fixed]    & $1.20$ [fixed]   & $1.20^{+0.01}_{-0.00}$    \\
$\alpha_{out}$ & $-15.27^{+0.15}_{-0.13}$  & $-15.76^{+0.17}_{-0.19}$  & $-16.35^{+0.09}_{-0.15}$  \\
\hline                                  
$\chi^2_r(ADI)$ &  0.927 &  0.886 & 1.211 \\
$\chi^2_r(PDI)$ &  1.493 &  1.095 & 1.652 \\
$\chi^2_r$      &  2.420 &  1.981 & 2.864 \\
\hline                                  
\end{tabular}
\tablefoot{The errors bars are shown at 3$\sigma$. The last three rows correspond to the reduced $\chi^2$ obtained for each wavelength. We also added $\chi^2_r(ADI)$ and $\chi^2_r(PDI)$ for a comparison of their respective values.}
\end{table}

Overall, the values we obtained for the morphological parameters of \HR{} are close or compatible within error bars to results from previous works on other SPHERE/ZIMPOL observations \citep{Milli_2019,Olofsson_2019} or GPI-J band observations \citep{Arriaga_2020,Chen_2020}. However, as we can see in Table \ref{table_morph_param}, the values we obtain in this work for all three wavelengths are relatively close, but not compatible within error bars. Regarding the eccentricity, $e$, the strong discrepancy between the ZIMPOL and IRDIS data could be linked to a better estimation of the star's position using IRDIS rather than ZIMPOL. However, for the semi-major axis $a$, the position angle, $PA$, the inclination, $i$, and the argument of the pericenter, $\omega$, we would expect the values to be compatible with one another. This may be explained by the extremely small error bars we are obtaining through the MCMC approach, which are most likely underestimated \citep{Mazoyer_2020}. \\
Regarding the outer slope, $\alpha_{out}$, the values we obtained reveal a relatively steep slope, which is consistent with the very narrow ring we observed. A variation depending on the wavelength could be expected since, at smaller wavelengths, we are probing smaller particles that are less bound than larger ones. This would result in a steeper slope for longer wavelengths, which we can tentatively see in our results (in particular, the IRDIS value compared to the ZIMPOL ones). In \citet{Kueny_2026}, the extreme adaptive optics instrument on the Magellan telescope (Mag-AO-X) was used, resulting in $\alpha_{out}$ values in Mag-AO-X [r'] ($\lambda \sim 615~nm$) and Mag-AO-X [i'] ($\lambda \sim 762~nm$) through two modeling methods: using \texttt{DiskFM} \citep{Mazoyer_2020} and a freeform model. The $\alpha_{out}$ values they obtain with their freeform model  are much bigger than with \texttt{DiskFM} (i.e. the freeform model shows a steeper slope). Interestingly, our ZIMPOL [R'] value is in between their two values ($|\alpha_{out,r'}|[\texttt{\small{DiskFM}}]<|\alpha_{out,R'}|[\texttt{\small{ZIMPOL}}]<|\alpha_{out,r'}|[\texttt{\small{Freeform}}]$), and our ZIMPOL [I'] value is very close to their \texttt{DiskFM} value  ($\alpha_{out,i'}[\texttt{\small{DiskFM}}]\sim\alpha_{out,i'}[\texttt{\small{ZIMPOL}}]<\alpha_{out,i'}[\texttt{\small{Freeform}}]$).\\
The value of the opening angle, $\Psi$, was left as a free parameter for the IRDIS modeling, but fixed for the ZIMPOL datasets (see the discussion in Appendix \ref{app_opang} on the difficulty to constrain $\Psi$). This value of $\Psi \sim 1.2\%$ is rather small compared to the value of 2.9\% in \citet{Olofsson_2022} or what could be expected from debris disks in general. In \citet{Thebault_2009} they find that for wavelengths where the smallest grains dominate the scattered light flux (i.e., in the visible to mid-infrared, MIR), a "natural" thickening happens even without any perturbing bodies, resulting in a minimum aspect ratio (equivalent to our definition of $\Psi$) value of $h_{min}\sim 4\pm 2\%$. The value we obtained is thus smaller than this "natural" limit proposed by \citet{Thebault_2009}; although recent results have suggested similar opening angles for \HR{} \citep{ARKS_III, Kueny_2026}. Such a small value could indicate a dynamically cold disk, with few collisions happening, which is surprising for such a young disk. This phenomenon, as well as the very steep inner slope, $\alpha_{in}$, could also be explained by planetesimal cores shepherding the dust observed in scattered light \citep{Lisse_2017} or a planet sculpting the inner edge of the disk \citep{Lagrange_2012} (although no planet has been observed in this system).

\subsection{Phase functions}
\label{results_phf}
The parametric approaches we chose for both the SPFs and the DoLPs allowed us to obtain a description of the phase functions over the whole scattering range (from 0 to 180°). However, the residual noise and self subtraction we obtained around the star (see Fig.\,\ref{fig_Obs_all} top row) makes it impossible to retrieve the correct phase function at small and big scattering angles \citep{Juillard_2023}. For this reason, all the following figures are shown between 13° and 167° (range of accessible angles in the IRDIS data). Additionally, as the ZIMPOL data are noisier than the IRDIS ones (particularly close to the star; see Appendix \ref{app_res_best_models}), the range is even smaller: from 35° to 145°. To illustrate this reduced range, the areas where the angles are not accessible for ZIMPOL (but still accessible for IRDIS) are shaded in grey. The error bars shown in Fig.\ref{fig_dhg_obs} and Fig.\ref{fig_dolp_obs} are obtained by drawing a large number of SPF realizations from the posterior distributions of our free parameters. For each scattering angle, we then extracted the 0.1th and 99.9th percentile of the SPFs which corresponds to the 3 $\sigma$ lower and upper bound.

\subsubsection{Scattering phase function}
\label{sec_spf_res}

\begin{figure}[ht!]
\centering
\includegraphics[width=\hsize]{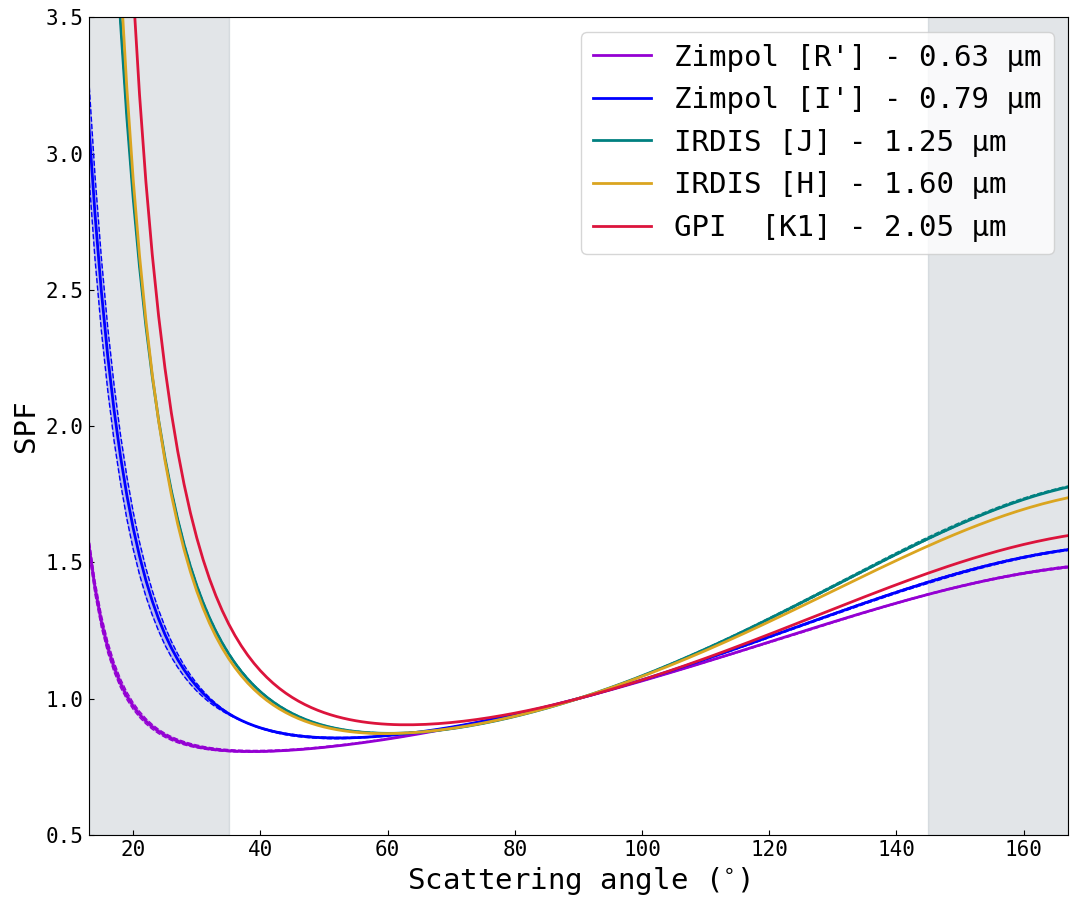}
    \caption{Double Henyey-Greenstein functions representing the scattering phase function of the best models for ZIMPOL R' (violet), ZIMPOL I' (blue), IRDIS (teal), with error bars shown at $3\sigma$. Data from \citet{Chen_2020} for IRDIS H band (yellow) and GPI K1 band (red) are also plotted, without the error bars. The shaded area corresponds to the scattering angles accessible in the IRDIS data but not in the ZIMPOL ones.}
    \label{fig_dhg_obs}
\end{figure}

Using the best models presented in \ref{sec_mcmc}, we plot the scattering phase function for each of the datasets in Fig.\ref{fig_dhg_obs}, with the SPFs normalized at 90° for better readability. The best-fit parameters obtained for $g_1$; $g_2$ and $w$ for our observations are presented in Table \ref{table_spf_param}, alongside the results from previous works: MagAO-X [g'], [r'], [i'], and [z'] bands \citep{Kueny_2026}, GPI [J] band, IRDIS [H] band, and GPI [K1] band \citep{Chen_2020}.  

\begin{table*}[ht!]
%\small
\fontsize{9.4pt}{9.4pt}\selectfont
\begin{center}    
\renewcommand{\arraystretch}{2} % Default value: 1
\caption{Phase function parameters obtained for the ZIMPOL and IRDIS models and comparison with other works.}  
\label{table_spf_param}   
%\centering                        
\begin{tabular}{c | c | c | c | c | c | c | c | c | c | c }  
\hline               
 & MagAO-X & MagAO-X & \underline{ZIMPOL}  & MagAO-X & \underline{ZIMPOL} & MagAO-X & \underline{IRDIS} & GPI & IRDIS & GPI\\         
 & [g']\tablefootmark{a} & [r']\tablefootmark{a} & [R'] &  [i']\tablefootmark{a} &[I']& [z']\tablefootmark{a} & [J]& [J]\tablefootmark{b} & [H]\tablefootmark{c} & [K1]\tablefootmark{b} \\   
\hline
$\lambda_c$ & 0.53 µm & 0.62 µm & 0.63 µm & 0.76 µm & 0.79 µm & 0.91 µm & 1.25 µm & 1.25 µm & 1.6 µm & 2.05 µm\\   
\hline                      
$g_1$  
& $0.80^{+0.01}_{-0.01}$ 
& $0.82^{+0.01}_{-0.01}$ 
& $0.97^{+0.01}_{-0.02}$    
& $0.91^{+0.02}_{-0.02}$ 
& $0.85^{+0.01}_{-0.01}$     
& $0.96^{+0.04}_{-0.05}$ 
& $0.86^{+0.01}_{-0.01}$ 
& $0.80^{+0.04}_{-0.04}$  
& $0.90^{+0.03}_{-0.03}$              
& $0.94^{+0.03}_{-0.04}$ \\ 
$-g_2$  
& $0.20^{+0.01}_{-0.01}$ 
& $0.20^{+0.00}_{-0.00}$ 
& $0.12^{+0.00}_{-0.00}$ 
& $0.15^{+0.00}_{-0.00}$ 
& $0.14^{+0.00}_{-0.00}$  
& $0.14^{+0.01}_{-0.01}$ 
& $0.18^{+0.01}_{-0.01}$ 
& $0.21^{+0.03}_{-0.03}$ 
& $0.17^{+0.01}_{-0.01}$             
& $0.15^{+0.01}_{-0.01}$ \\
$w$    
& $0.29^{+0.01}_{-0.01}$ 
& $0.30^{+0.01}_{-0.01}$ 
& $0.14^{+0.05}_{-0.05}$  
& $0.37^{+0.04}_{-0.04}$ 
& $0.13^{+0.01}_{-0.01}$           
& $0.52^{+0.4}_{-0.2}$ 
& $0.26^{+0.01}_{-0.01}$  
& $0.34^{+0.03}_{-0.03}$  
& $0.32^{+0.07}_{-0.04}$              
& $0.5^{+0.2}_{-0.1}$  \\
\hline     
\end{tabular}
\end{center}
\tablefoot{For the data obtained in this work (ZIMPOL [R'], ZIMPOL [I'], IRDIS [J], underlined in the table), the error bars are shown at 3$\sigma$. The error bars obtained from \citet{Chen_2020} and \citet{Kueny_2026} are shown at 1$\sigma$. \\
\tablefoottext{a}{Mag-AO-X data obtained by \citet{Kueny_2026}.} 
\tablefoottext{b}{GPI [J] and GPI [K1] data obtained by \citet{Chen_2020}.} 
\tablefoottext{c}{IRDIS [H] data obtained by \citet{Chen_2020} by re-analyzing the data presented in \citet{Milli_2017}.} }
\end{table*}

For all the three cases presented in this work, the results show a strongly forward scattering SPF, with a high value of $g_1$. Our results in the J-band are close to, but not compatible with the GPI [J] data obtained by \citet{Chen_2020}, while our ZIMPOL [R'] and [I'] are not compatible with values from \citet{Kueny_2026} either. This may result from our very small error bars, which are most likely underestimated, but it is also interesting to note that \citet{Kueny_2026} reported that the SPF they obtained from their freeform approach is notably distinct from the one they obtained using a double Henyey-Greenstein parametrization.
When considering the variation of the three parameters as a function of $\lambda$ (Table \ref{table_spf_param}), no clear trend can be observed in our data, nor when including data from previous works at other wavelengths. Interestingly, at small scattering angles, the variation in the SPF as a function of the wavelength observed in Fig.\,\ref{fig_dhg_obs} is similar to that observed by \citet{Chen_2020} in their DHS modeling of \HR. Their best model consisted in a mixture of amorphous silicates (42 vol.\%), amorphous carbon (17 vol.\%) and metallic iron (37 vol.\%), with a minimum particle size of $s\sim25$ µm.

\subsubsection{Degree of linear polarization}
\label{sec_dolp_res}
\begin{figure}[ht!]
\centering
\includegraphics[width=\hsize]{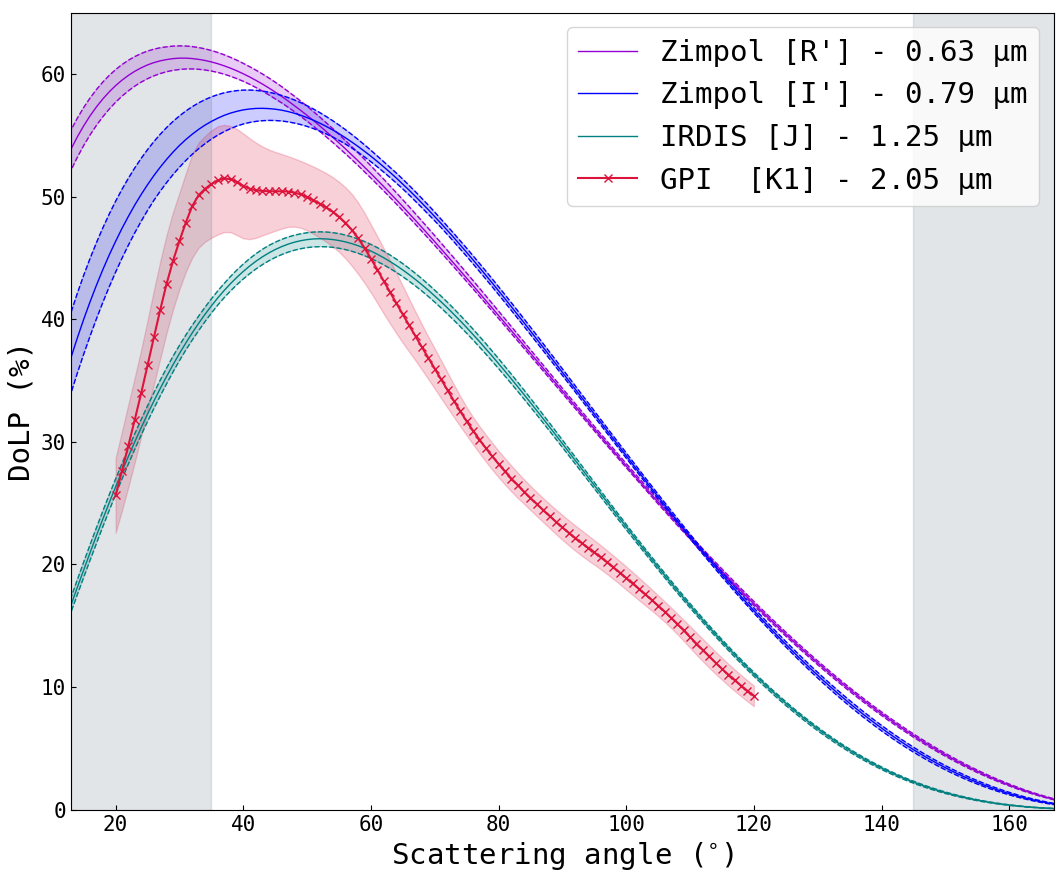}
    \caption{Beta functions parametrizing the degree of linear polarization of the best models for ZIMPOL R' (violet), ZIMPOL I' (blue), IRDIS (teal), with the error bars shown at 3$\sigma$. The DoLP curve (non parametric fit) obtained by \citet{Arriaga_2020} is also plotted (red). The shaded area corresponds to the scattering angles accessible in the IRDIS data but not the ZIMPOL ones.}
    \label{fig_dolp_obs}
\end{figure}

Fig. \ref{fig_dolp_obs} shows the degree of linear polarization we obtained with our parametrization for ZIMPOL R' (violet), ZIMPOL I' (blue) and IRDIS J (teal). To increase the number of studied wavelengths, we added the DoLP obtained by \citet{Arriaga_2020} for GPI K1. Their approach is different, as they did not parametrize the DoLP; instead, they obtained it by modeling the SPF and pSPF in a similar way to the approach presented in Appendix \ref{param_descr_phf}. The parameters $\alpha$, $\beta$, and $f_M$ obtained in our parametric approach are described in Appendix \ref{app_param_beta_func}.

From Fig. \ref{fig_dolp_obs}, we can also observe that the DoLP is wavelength dependent, although without any clear trends. When considering only our parametric approach, (R', I', and J bands); there seems to be a tendency where the longer the wavelength, the lower the value of the degree of linear polarization, and the smaller the peak scattering angle. Nonetheless, this is not true over the whole range of scattering angles: the DoLP of ZIMPOL R' (violet) becomes smaller than the one of ZIMPOL I' (blue) between 50° and 110°. 
Additionally, when considering the DoLP found by \citet{Arriaga_2020} at 2.05 µm, the trend reverses, as the peak value is higher than the one obtained for IRDIS J (1.25 µm) and for smaller scattering angles. However, as previously mentioned, the error bar we obtained from the MCMC approach are most likely underestimated, meaning that we cannot conclude on any trend. Another approach to constrain the variation of the degree of linear polarization as a function of the wavelength is presented below in \ref{subsec_90}.\\
Table \ref{table_doLP_obs_comp} compares the maximum DoLP value $P_M$ and the corresponding scattering angle $\theta_M$ for \HR{} at several wavelengths and for other debris disks. Overall \HR{} shows a strong polarization peak value, $P_M$, at small scattering angles ,$\theta_M$, in particular when compared to other debris disks. Possible explanations for such a high degree of linear polarization are discussed in Sect.\,\ref{sec_discussion}. 

\begin{table}[ht!]
\renewcommand{\arraystretch}{2} 
\caption{Maximum value $P_M$ and scattering angle $\theta_M$ of the DoLP peak for \HR{} (this work) and comparison with other debris disks.}                
\label{table_doLP_obs_comp}   
\centering                        
\begin{tabular}{c | c | c | c  }      
\hline        
Disk  & Band & $P_M$ & $\theta_M$ \\
\hline
\HR{} & R' & $\sim$61 \% & $\leq$ 35° \\
\hline
\HR{} & I' & $\sim$57 \% & $\sim$40° \\
\hline
\HR{} & J & $\sim$47 \% & $\sim$50° \\
\hline
HD\,181327 \footnotesize{[\citealp{Milli_2024}]} & H & $\sim$23 \% & $\sim$80° \\
\hline
HD\,114082 \footnotesize{[\citealp{Engler_2023}]} & H & $\sim$12 \% & $\sim$90° \\
\hline
HD\,35841 \footnotesize{[\citealp{Esposito_2018}]}  & H & $\sim$30 \% & $\sim$100° \\
\hline                                  
\end{tabular}
\end{table}

As stated in \citet{Chen_2020}, their DHS model shows a polarization fraction close to 100\%, which is incompatible with the observations presented in \citet{Milli_2019} or \citet{Arriaga_2020}. Such a high DoLP is also incompatible with our observations, but, their overall shape is similar than our result, with a low $\theta_M$, such that $\theta_M~\leq$ 60°. Additionally, the variation of the DoLP with the wavelength in their model shows the same trend than what we obtain in the ansae: the longer the wavelength, the lower the DoLP value at 90° (see Sect.\,\ref{subsec_90}).

\subsection{Degree of linear polarization and spectral reflectance in the ansae}
\label{subsec_90}
\begin{figure}[ht!]
\centering
\includegraphics[width=\hsize]{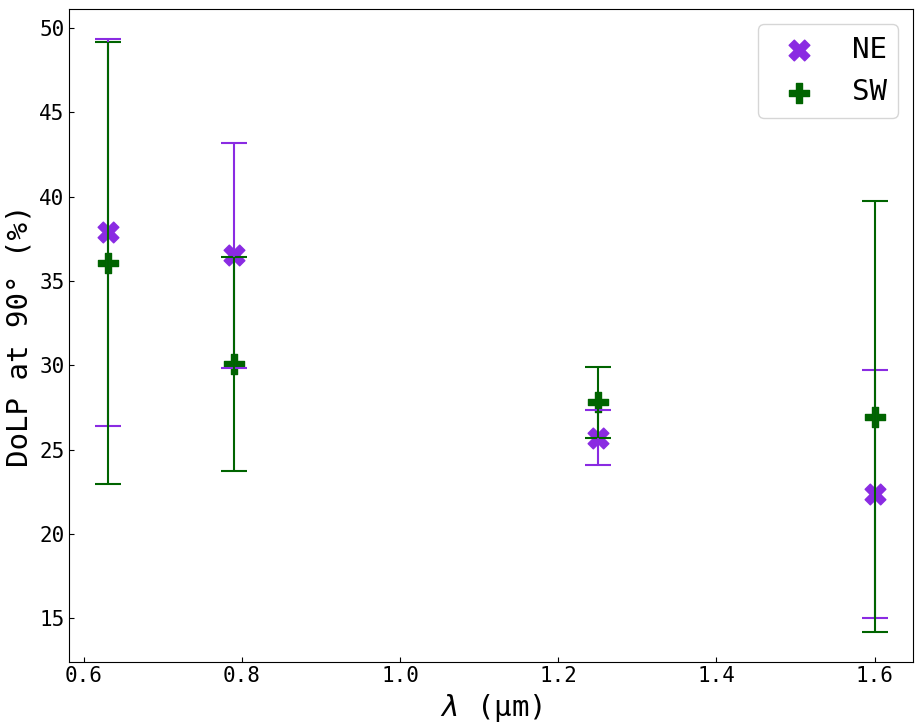}
\includegraphics[width=\hsize]{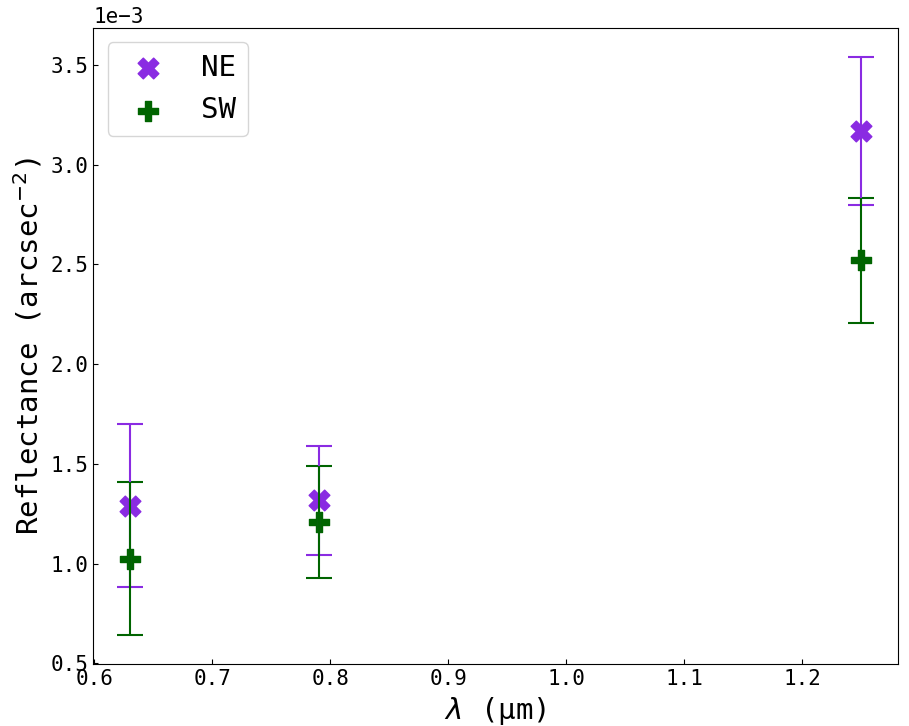}
    \caption{DoLP in \% (top) and reflectance in contrast per arcsec$^{2}$ (bottom) obtained on our observations. The northern ansae is represented in violet crosses, and the southern ansae in green plusses. The three wavelength we studied in this work are plotted: 0.63 µm (ZIMPOL [R']); 0.79 µm (ZIMPOL [I']); 1.25 µm (IRDIS [J]) for both the DoLP and the spectral reflectance. The DoLPs we obtained for each ansae using the data from \citet{Crotts_2025} for 1.60 µm (GPI) are also shown.}
    \label{DoLP_Refl_90}
\end{figure}

Due to the stellar flux residuals around the coronagraph, the disk is best constrained in the ansae, where the scattering angle is about 90°. To have a complementary nonparametric and nonsymmetric approach to characterize our disk, we decided to perform aperture photometry on all our data. We extracted both the spectral reflectance and the degree of linear polarization in the ansae at a 90° scattering angle. For the degree of linear polarization, we also used data from GPI in the H band\footnote{data available at \url{https://www.canfar.net/citation/landing?doi=24.0089}} (1.60 µm) \citep{Crotts_2025}. The process is detailed in Appendix \ref{app_aper_phot}. The obtained degrees of linear polarization and spectral reflectance for each ansa are plotted in Fig. \ref{DoLP_Refl_90} (top and bottom), with the values presented in Table \ref{tab_dolp90} and Table \ref{tab_refl90}, respectively. Also, NE refers to the northeast ansa and SW to the southwest ansa.

\begin{table}[ht!]
\renewcommand{\arraystretch}{2} 
\caption{Degrees of linear polarization measured on the observational data for a 90° scattering angle.}     
\label{tab_dolp90}   
\centering                        
\begin{tabular}{c | c | c | c | c }      
\hline               
 Band & R' & I' & J & H  \\         
\hline
$\lambda_c$ (µm)& 0.63 & 0.79& 1.25 & 1.60 \\         
\hline 
NE (\%) & 37.9 $\pm$ 11.5 & 36.5 $\pm$ 6.7 & 25.7 $\pm$ 1.6 & 22.4 $\pm$ 7.4 \\
SW (\%) & 36.1 $\pm$ 13.1 & 30.1 $\pm$ 6.4 & 27.8 $\pm$ 2.1 & 27.0$ \pm$ 12.8\\
\hline                                  
Avg (\%)& 37.0 $\pm$ 8.7 & 33.3 $\pm$ 4.6 & 26.8 $\pm$ 1.3 & 24.7 $\pm$ 7.4  \\
\hline 
\end{tabular}
\end{table}

\begin{table}[ht!]
\renewcommand{\arraystretch}{2} 
\caption{Spectral reflectance in contrast /arcsec$^2$ measured on the observational data for a 90° scattering angle.}
\label{tab_refl90}   
\centering                        
\begin{tabular}{c | c | c | c  }      
\hline               
 Band & R' & I' & J   \\         
\hline
$\lambda_c$ (µm)& 0.63 & 0.79& 1.25  \\         
\hline 
NE (arcsec$^{-2}$)x$10^{-3}$& 1.3 $\pm$ 0.4 & 1.3 $\pm$ 0.3 & 3.2 $\pm$ 0.4 \\
SW (arcsec$^{-2}$)x$10^{-3}$& 1.0 $\pm$ 0.4 & 1.2 $\pm$ 0.3 & 2.5 $\pm$ 0.3 \\
\hline                                  
Avg(arcsec$^{-2}$)x$10^{-3}$& 1.2 $\pm$ 0.4 & 1.3 $\pm$ 0.3 & 2.9 $\pm$ 0.3 \\
\hline 
\end{tabular}
\end{table}

Within the error bars, the degree of linear polarization and the spectral reflectance are compatible between the two ansae. The reflectance seems higher in the NE than in the SW, but not in a significant way. This could tentatively be an indication of a different dust population between the two ansae, but it will not be studied further here. We also plan to consider the averaged value between the NE and SW ansae in later discussions. The reflectance additionally shows a red spectral slope, as was previously observed for this disk \citep{Rodigas_2015, Milli_2017}. \\
The DoLP at 90° we obtained is slightly different from what we obtained with our parametric description (beta function). Overall, the averaged DoLP value we obtained through aperture photometry is higher in the visible (but still compatible within error bars). It is lower for IRDIS and compatible only with the results from the SW ansa. Additionally, the DoLP at 90° seems to decrease with the wavelength consistently: at $\lambda$ = 0.79 µm, it is lower than at $\lambda$ = 0.63 µm (which was not the case in our models). This may result from our bell-shape parametric description not being perfectly adapted to describe \HR's DoLP. In \citet{Ren_2023}, they used this three-parameter bell shape to fit DoLPs peaking around 90°, a typical $\theta_M$ for astronomical and solar system object, and at low $P_M$, whereas \HR{} shows a high $P_M$ and low $\theta_M$, in particular in the visible.

\section{Laboratory measurements}
\label{sec_lab_mes}
To help constrain the physicochemical properties of the dust particles orbiting in \HR, we experimentally measured the optical properties of a chosen sample. We then compared those values to our observations. 

\subsection{Choice of the sample}
\label{subsec_choice_sample}
As previously mentioned, the DoLP of \HR{} peaks at higher values and smaller scattering angles than other debris disks observed so far \citep{Milli_2024, Engler_2023, Esposito_2018}. To reproduce these peculiar values, the sample needs to be highly polarizing, and thus very absorbent \citep{Dollfus_1971}. Following previous results \citep{Milli_2023} and experimental measurements available on the \texttt{PROGRA$^2$} database\footnote{\href{https://www.icare.univ-lille.fr/progra2-en/?noredirect=en_US}{\url{https://www.icare.univ-lille.fr/progra2-en/?noredirect=en_US}}} \citep{Worms_1999,Hadamcik_2009,Renard_2014}, we chose an iron sulfide sample, as these minerals have a low albedo (\citealp{deBergh_2008} and references therein) and a degree of linear polarization that peaks at high values for low scattering angles. Additionally, there is evidence to support the presence of such iron-bearing opaque minerals in interstellar molecular clouds and cores \citep{Pollack_1994}, and in small bodies of the Solar System. Iron sulfides are part of the (sub)micrometric mineral grains constituting the dark matrix of carbonaceous chondrites meteorites \citep{McSween_2010}, the Stardust samples showed the presence of Fe-Ni sulfides, dominated by troilite (FeS), in the comet 81P/Wild 2 \citep{Dobrica_2009}; and such Fe-bearing opaque minerals are also good candidates to reproduce the low albedo of the comet 67P/Churyumov-Gerasimenko \citep{Capaccioni_2015,Quirico_2016,Rousseau_2018}. 
Moreover, Fe–Ni sulfides are the second most common minerals in anhydrous chondritic porous interplanetary dust particles (CP-IDPs) after crystalline silicates \citep{Dai_2001,Bradley_2014}. 
Finally, submicrometric Fe-bearing opaque mineral grains seem to be the main reason for the low albedo in the visible to MIR of comets, dark asteroids (B,C,D,P,Z-types), and carbonaceous chondrites, whereas silicates and organic matter are expected to contribute less to their optical properties at these wavelengths \citep{Quirico_2016,Rousseau_2018,Beck_2025}. \\
Our iron sulfide sample is composed of compact irregular particles, composed of 55 vol\% of troilite FeS and at 45 vol\% of pyrrhotite Fe\textsubscript{1-x}S (with 0 < x < 0.2). It was characterized by scanning electron microscopy (see Fig.\ref{Fig_FeS_MEB}) and the size distribution of the particles was measured from these SEM images (see Fig.\ref{fig_size_distr_FeS}). From this measurement, all particles were found to be smaller than 100 µm, mostly larger than 1 µm, and with an average size between 5 and 10 µm. This sample will further be referred to as "s < 100 µm". For a comparison, we also measured the powder of the same sample material obtained after grinding down to particles smaller than 1 µm. This second sample is referred to as "s < 1 µm", has an average particle size $\sim$ 0.3 µm \citep{Sultana_2023}, and SEM images of this powder are shown in Fig.\,\ref{Fig_FeS_MEB_small}.
The choice of a purely opaque mineral sample was motivated by the unusual characteristics of \HR's DoLP; however, it is important to note that the detailed composition of the disk is most likely much more complex and a mixture of various components, such as astrosilicates, carbonaceous compounds, and so on, as can be found in comets, CP-IDPs, or carbonaceous chondrites. 

\subsection{Experimental setup}
\label{subsec_exp_setups}
We used the Spectro-photometer with cHanging Angles for Detection Of Weak Signals (\texttt{SHADOWS}\footnote{\href{https://cold-spectro.sshade.eu/-SHADOWS-Micro-Spectro-Gonio-Radiometer-}{\url{https://cold-spectro.sshade.eu/-SHADOWS-Micro-Spectro-Gonio-Radiometer-}}}) spectro-gonio-radiometer \citep{Potin_2018} to measure the DoLP of our sample (more accurately the DoLP\textsubscript{Q}. More details are given in Appendix \ref{app_calib_shadows} and especially Eqs. (\ref{eqDoLP}) and (\ref{eqDoLPQ})). A monochromatic unpolarized light (DoLP < 2.3 \%) illuminates at a controlled incidence angle the surface of a mineral powder deposited horizontally, and the reflected light is measured at a controlled emergence angle by detectors placed after polarizers (see Appendix \ref{exp_setup} for a detailed description of the experimental setup and Appendix \ref{app_calib_shadows} for the calibration and uncertainty sources of the setup).\\ 
The main caveat with this experiment is that the particles are deposited rather than in suspension, which may change their optical properties. However, as FeS is a very absorbing material, multiple scattering is null or negligible, and the polarization is governed by single-scattering for deposited as well as for lifted particles \citep{Hadamcik_2023}. Therefore, for absorbing particles, the differences in optical properties between deposited and lifted states should be lessened. To investigate this hypothesis on our samples, we used the \texttt{PROGRA$^2$} database to compare our measurements on deposited FeS samples (using \texttt{SHADOWS}) to those on lifted FeS samples (using \texttt{PROGRA2}). \texttt{PROGRA$^2$} allows us to measure the DoLP of samples either in micro-gravity or lifted through an air draught. These results are shown in Fig.\,\ref{fig_DoLP_Mes} and Table \ref{tab_comp_liff}.

\subsection{Results}
\label{subsec_res_labmes}
\begin{figure*}[ht!]
\centering
\includegraphics[width=1\textwidth]{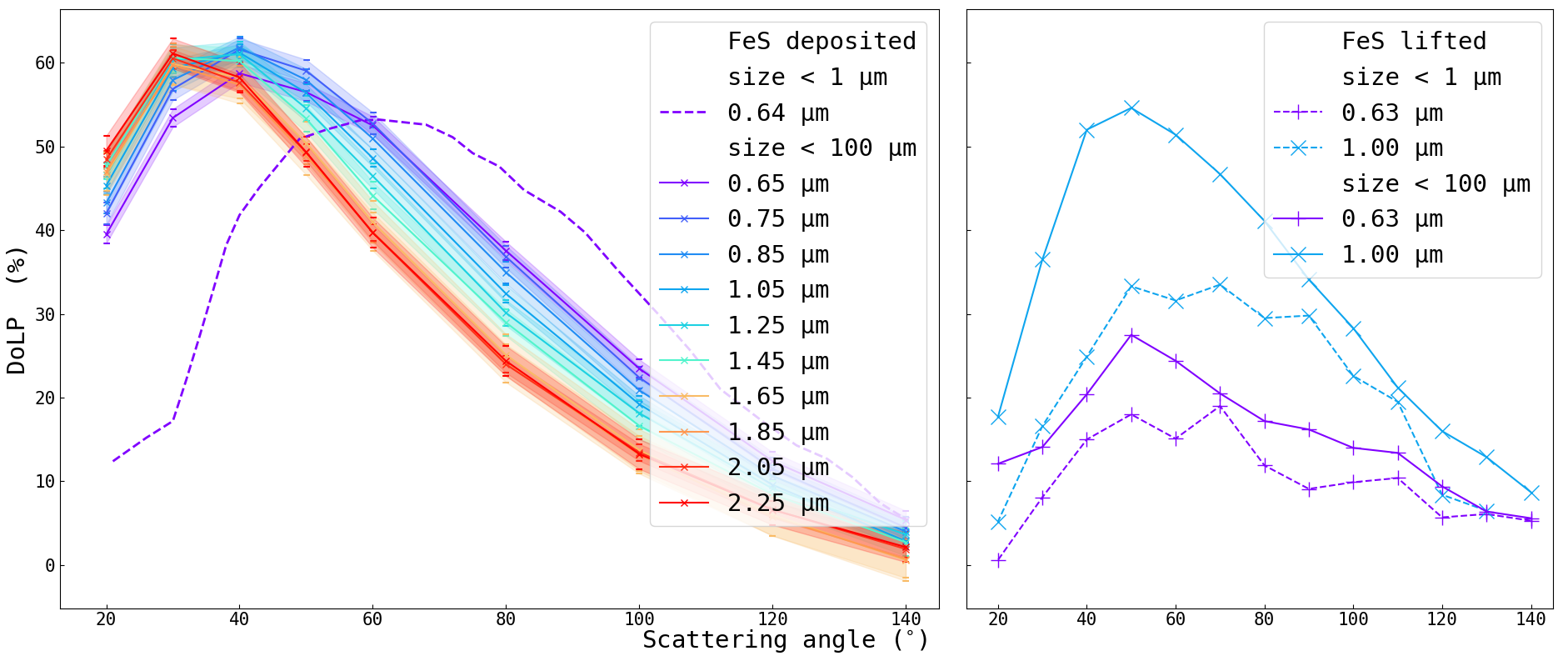}
    \caption{Experimentally measured DoLP of two samples (one < 100 µm, and another < 1 µm) made of the same iron sulfide mixture (55 vol\% troilite and 45 vol\% pyrrhotite), labeled "FeS."
    Left: Deposited FeS sample, measured with the \href{https://cold-spectro.sshade.eu/-SHADOWS-Micro-Spectro-Gonio-Radiometer-}{\texttt{SHADOWS}} spectro-goniometer. The black dotted line is the DoLP for a sample with small particles (size $s<$ 1 µm), taken at 0.64 µm, whereas the rainbow colours represent various wavelengths (from 0.65 to 2.25 µm) for a sample with particles of size $s < 100$ µm.
    Right: lifted FeS sample, measured with the \href{https://www.icare.univ-lille.fr/progra2-en/?noredirect=en_US}{\texttt{PROGRA$^2$}} experiment. The violet curves are the measurements taken at 0.63 µm, whereas the blue curve were taken at 1.00 µm. The dotted lines are the DoLP for the sample with particles size < 1 µm, taken using an air draught and the plain lines represent a sample with $s < 100$ µm, measured in microgravity.}
    \label{fig_DoLP_Mes}
\end{figure*}

Fig.\ref{fig_DoLP_Mes} left presents polarimetric phase curves  measured with \texttt{SHADOWS} on the FeS sample with a size < 100 µm, generated from $\lambda$ = 0.65 µm to $\lambda$ = 2.95 µm, with a step of 0.1 µm; for scattering angles varying from 20° to 140°. Overplotted in dashed violet line, we have the DoLP of the sample with submicrometer size particles (s < 1 µm) at $\lambda$ = 0.64 µm for comparison.\\
Fig.\ref{fig_DoLP_Mes} right shows the DoLP at $\lambda$ = 0.63 µm and $\lambda$ = 1.00 µm of the FeS sample (s < 100 µm), obtained through the \texttt{PROGRA$^2$} database. The FeS sample (size < 1 µm) was measured using an air draught, while the DoLP of the sample with large particles (s < 100 µm) was obtained in microgravity. \\
Table \ref{tab_comp_liff} extracts the DoLP peak value $P_M$ and peak scattering angle, $\theta_M$, of the various measurements presented in Fig.\,\ref{fig_DoLP_Mes}. 

\begin{table}[ht!]
\renewcommand{\arraystretch}{2} 
\caption{Comparison of the $P_M$ and $\theta_M$ for deposited and lifted particles.}                
\label{tab_comp_liff}   
\centering                        
  \begin{tabular}{c | c | c | c | c }
    \hline
    $s < 100$ µm & \multicolumn{2}{c|}{Deposited} & \multicolumn{2}{c}{Lifted}\\
     \hline
        & \multicolumn{2}{c|}{SHADOWS} & \multicolumn{2}{c}{PROGRA$^2$ (µ-grav)}\\
    \hline
     $\lambda$ & 0.65 µm & 1.05 µm & 0.63 µm & 1.0 µm\\
    \hline
       $P_M$   & 58.7\% & 61.2\% & 27.5\% & 54.6 \% \\ 
    \hline
    $\theta_M$ & 40° & 40° & 50° & 50° \\ 
    \hline
    \hline
     $s < 1$ µm& \multicolumn{2}{c|}{Deposited} & \multicolumn{2}{c}{Lifted}\\
    \hline
                & \multicolumn{2}{c|}{SHADOWS} & \multicolumn{2}{c}{PROGRA$^2$ (a.d.)}\\
     \hline
     $\lambda$ & \multicolumn{2}{c|}{0.64 µm} & 0.63 µm & 1.0 µm\\
    \hline
       $P_M$   & \multicolumn{2}{c|}{53.2\%} & 19\% & 33.5\% \\ \hline
    $\theta_M$ & \multicolumn{2}{c|}{61.0°}  & 70°  & 70°    \\ \hline
  \end{tabular}
\tablefoot{The first half of the table corresponds to our main sample ($s < 100$ µm), and the second half to an another FeS sample with smaller particles ($s < 1$ µm). In the case of the large particles, the sample was lifted in microgravity (µ-grav), whereas the small particle sample was lifted through an air draught (a.d.).}
\end{table}

The main result obtained from Fig.\,\ref{fig_DoLP_Mes} and Table \ref{tab_comp_liff} is that when considering either lifted or deposited particles, $\theta_M$ is 20° smaller for FeS (s < 100 µm) than $\theta_M$ for FeS (s < 1 µm). Additionally, for large particles (s < 100 µm) in the near-IR, the $P_M$ values of deposited and lifted particles are both high and very similar. \\
Tables \ref{table_doLP_obs_comp} and \ref{tab_comp_liff} show that the very small $\theta_M$ obtained in our observations of \HR{} ($\theta_M \leq$ 50°) is well reproduced by our large particles (s < 100 µm) FeS sample (both when lifted and deposited). Comparing Fig.\,\ref{fig_DoLP_Mes} with Fig.\,\ref{fig_dolp_obs} also shows that this sample reproduces well the high $P_M$ value of the disk, as well as its overall DoLP shape, whereas FeS (s < 1 µm) does not. Overall, these results are arguing in favour of large opaque mineral particles being the main scatterers in \HR.\\
It is interesting to note that (for the same particle size) we are seeing that $\theta_M(dep)$ is shifted of $\sim$-10° compared to $\theta_M(lift)$, as observed in \citet{Hadamcik_2023}. However, there are discrepancies between $P_M(dep)$ and $P_M(lift.)$ in the visible since, based on the work by \citet{Hadamcik_2023}, we would expect the $P_M$ value to be similar for deposited and lifted samples at a given particle size. There is also a surprisingly significant difference for lifted large (s < 100 µm) particles in the visible and NIR. There are several possible explanations for these effects: the small particles may have been altered, the visible and IR detectors do not have the same sensitivity in the \texttt{PROGRA$^2$} experiments, the small particles were lifted with an air draught (whose efficiency depends on the particle size) whereas the large ones were measured in micro-gravity, etc. However, we did not investigate this issue further, as it is beyond the scope of this paper.\\
Fig.\,\ref{fig_DoLP_Mes} left shows the variation of the DoLP curve as a function of the wavelength. The DoLP peak value $P_M$ of the FeS (s < 100 µm) sample is around 60\% for every wavelength, and increasing the wavelength seems to diminish $\theta_M$: for $\lambda$ = 2.25 µm, $\theta_M \sim$ 30°, against 40° at $\lambda$ = 0.65 µm. Finally, for $\theta~\in$ [50°,140°], there is a clear wavelength dependency: the lower the wavelength, the higher $P_M$ is.

\section{Discussion and summary}
\label{sec_discussion}
\subsection{Large opaque minerals as the main scatterers in \HR}
As discussed in Sect.\,\ref{subsec_res_labmes}, the FeS (s < 100 µm) sample provides a good match for the polarimetric characteristics of \HR. To add to this comparison, we further studied the DoLPs at 90° as a function of the wavelength of our observations and measurements, as well as their spectral reflectance.

\subsubsection{DoLP at 90°}
\label{subsec_DoLP_Res}
\begin{figure}[ht!]
\centering
\includegraphics[width=\hsize]{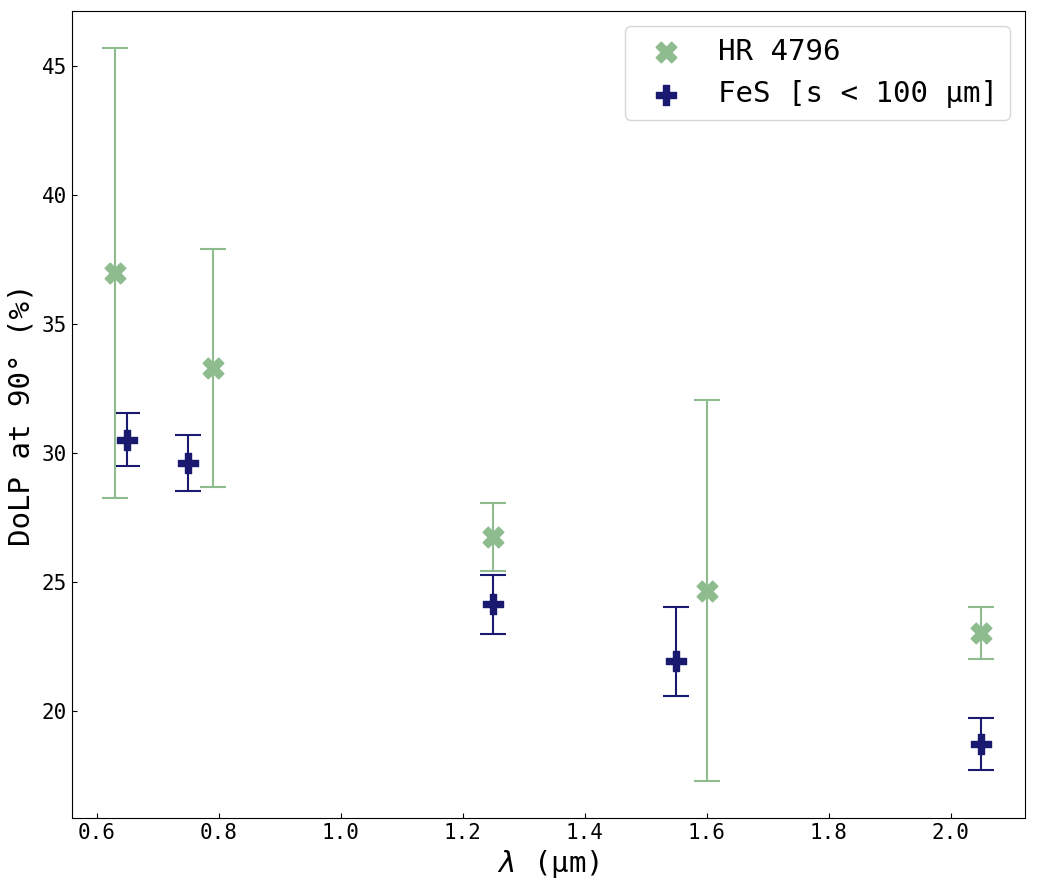}
    \caption{DoLP values as a function of the wavelength, for both the observational data (dark blue plus symbols) and the experimental measurements on the large particle (s < 100 µm) iron sulfide sample (light green crosses). The GPI-K data (HR 4796 at $\lambda$ = 2.05 µm) are from \citet{Arriaga_2020}.}
    \label{DoLP_90_All}
\end{figure}

As previously mentioned, our parametric approach for the beta function might not be perfectly suited to describe \HR's DoLP. To avoid any parametrization, in Fig.\,\ref{DoLP_90_All}, we compare the DoLPs at 90° of \HR{} obtained through aperture photometry in the ansae (see Sect.\,\ref{subsec_90} and Appendix \,\ref{app_aper_phot}) for ZIMPOL R', ZIMPOL I', IRDIS, and GPI-H with the DoLPs at 90° of our measurements for similar wavelengths. We also added the GPI-K DoLP value extracted from \citet{Arriaga_2020}. Although this value was not obtained through aperture photometry, their model was not parametric and considered brightnesses at specific scattering angles.

Figure \ref{DoLP_90_All} shows that at 90°, the DoLP we obtained for ZIMPOL R', ZIMPOL I', IRDIS J, and GPI H corresponds (within the error bars) to the ones we obtained for iron sulfide at similar wavelengths (0.65 µm, 0.75 µm, 1.25 µm, 1.55 µm, and 2.05 µm). At all wavelengths, for both the observations and the experimental measurements, the lower the wavelength, the higher the DoLP value at 90°.
Moreover, a negative slope of the degree of linear polarization, as seemingly observed in \HR, is a behaviour that is compatible with compact particles, of a size that is bigger than the observation wavelength \citep{Hadamcik_2003}; which would be bigger than a micron in our case.

\subsubsection{Spectral reflectance}
\label{subsec_Refl_Res}
Utilizing the results presented in Sect.\,\ref{subsec_90}, we were able to plot the spectral reflectance averaged over the two ansae for our data (ZIMPOL R', I', and IRDIS J), superimposed with previous measurements. We used the data \citet{Milli_2017} obtained by performing aperture photometry with a radius of 63 mas, and the results from the aperture photometry achieved on HST and near-IR Mag-AO data by \citet{Rodigas_2015} and visible Mag-AO-X by \citet{Kueny_2026}, with the side of the aperture being 152 mas (HST/STIS), 162 mas (HST/NICMOS), 111 mas (near-IR Mag-AO), and 63 mas (vis-Mag-AO-X). It must be noted that while the apertures were taken in the ansae for our measurements as well as those from HST and Mag-AO, the value from simultaneous IRDIS dual band and Integral Field Spectrometer (IFS) observations (IRDIFS) was averaged over the whole disk, excluding the ansae. We also added the spectral reflectance measured on iron sulfide samples of various sizes (50-100 µm; 25-50 µm; and 0.3 µm) with the \texttt{SHADOWS} instrument in \citet{Sultana_2022}. These measurements are dimensionless (normalized at 1 µm) and superimposed to the observations in such a way that we are able to compare the overall shapes of the curves (although not the values).

\begin{figure}[ht!]
\centering
\includegraphics[width=\hsize]{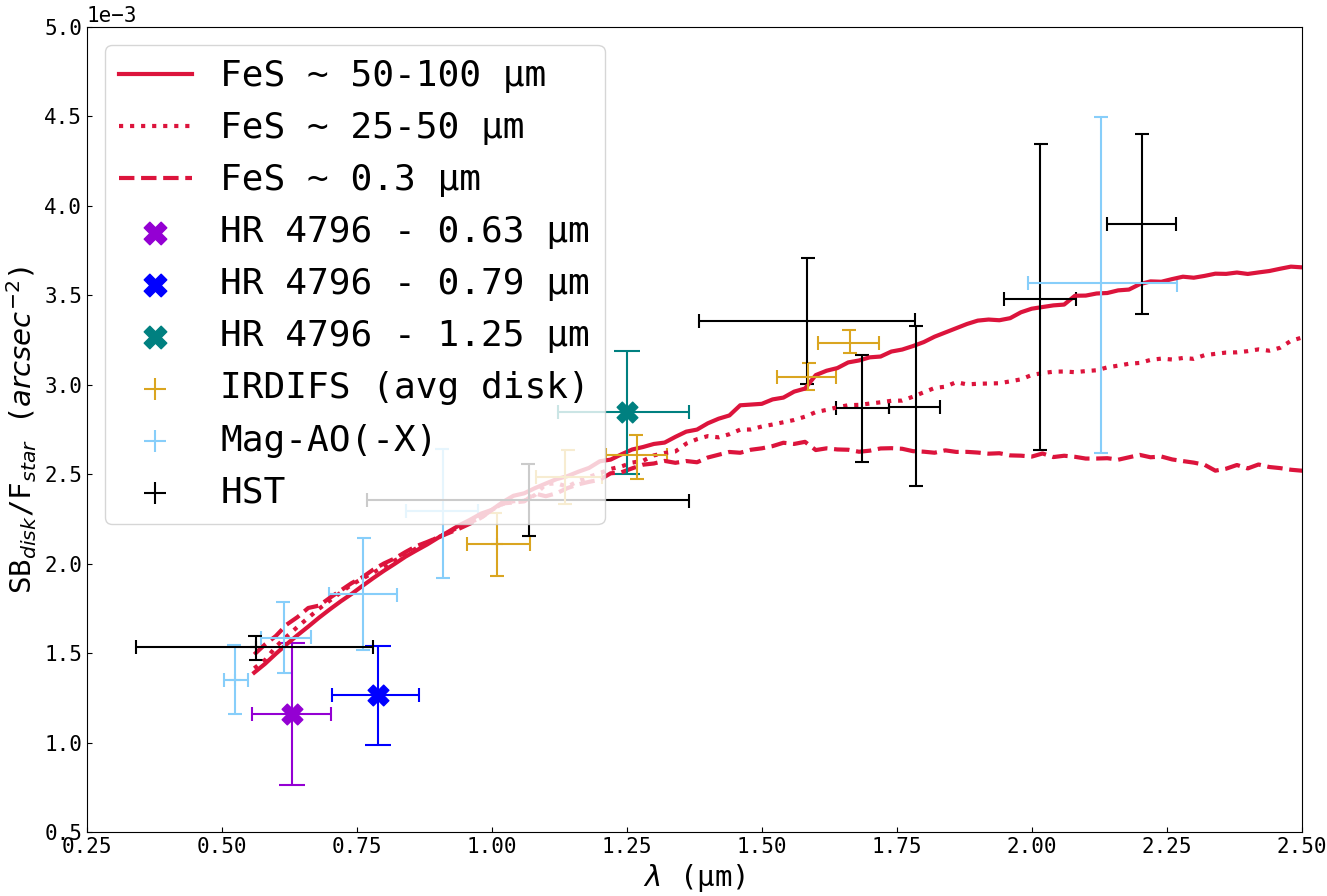}
    \caption{Normalized reflectance values as a function of wavelength for both observations of \HR{} (various crosses) and normalized experimental measurements on iron sulfide samples (crimson lines) from \citep{Sultana_2022}. The observational data are issued from this work (R' in violet, I' in blue, J in teal) and completed by data from IRDIFS in gold \citep{Milli_2017}, HST in black, Mag-AO \citep{Rodigas_2015}, and Mag-AO-X \citep{Kueny_2026} in light blue. While the photometry in this work as well as for the HST and Mag-AO data were all taken at 90° (in the ansae), the IRDIFS data are averaged over the whole disk, with the ansae removed. The horizontal error bars correspond to the FWHM of the filters used for each of the observation. The measured iron sulfides particles have different sizes, $s$: as a plain line, we have $s\sim$50-100 µm; as a dotted line, $s\sim$25-50 µm, and as a dashed line, $s\sim$0.3 µm.}
    \label{Refl_90_All}
\end{figure}

The spectral reflectances of the observations presented in this work are in agreement with the red slope previously observed on \HR{} over the visible to NIR range \citep{Debes_2008,Rodigas_2015,Milli_2017,Lisse_2017,Kueny_2026}. The value we obtained for ZIMPOL I' is slightly lower than what could be expected when comparing to previous measurements, but compatible within the error bars to the HST data \citep{Rodigas_2015} and recent Mag-AO-X results \citep{Kueny_2026}. \\
These results are also in agreement with the shape of the spectral slope measured for large iron sulfide particles ($s\sim$ 25-100 µm), which is steeper than for submicronic particles. Additionally, the gentler slope observed in the reflectance of large particles for longer wavelengths (around $\lambda$ = 2 µm) is also in agreement with a flattening of the slope observed by \citet{Rodigas_2015} at those wavelengths. 
\\
\\
From both the DoLPs at 90° and the spectral reflectance, large iron sulfides particles with a minimum size of a few micrometers to $ s < 100$ µm are still considered a good match to explain the peak $\geq$ 47 \% at a scattering angle $\leq$ 50° of \HR. Such a result would indicate that large opaque minerals dominate the scattering polarimetric properties of the disk. This result is in agreement with the conclusions presented in \citet{Kueny_2026}, where they found that their SPF demonstrates a match with large, highly absorbing particles of several micrometers in size.

\subsection{Comparison to Solar System objects}
\label{sec_comp_SS}
The main asteroid belt and the Kuiper belt, analogues to debris disks, are remnants of the formation of the Solar System. We have access to observations and in-situ measurements of the planetesimals (asteroids, comets, etc.) orbiting in them. In debris disks, we (mostly) observe the dust particles resulting from the collisions of such planetesimals and these particles serve as an indicator of the surface composition of the bodies in the disk. \\
Similarities between debris disks and Solar System objects have already been observed \citep{Engler_2023, Milli_2024, Lisse_2025}. In \citet{Milli_2024}, the authors reported that the averaged DoLP value for comets in the NIR obtained by \citet{Kiselev_2015}, offers a relatively good match to the DoLP they obtained for HD~181327. This result is in agreement with the results from \citet{Xie_2025}, where, similarly to Kuiper Belt Objects in the Solar System, they found water ice in HD~181327. \newline
As seen in Table \ref{table_doLP_obs_comp}, the \HR{} peaks for much smaller scattering angles and at higher values than HD~181327. As previously discussed, these characteristics can be reproduced by dark samples \citep{Dollfus_1971}, which is why we looked for possible comparisons with low albedo objects in the Solar System. For example, C-type asteroids represent the main population of the asteroid belt \citep{Noguchi_2023}and usually present a low albedo \citep{Vernazza_2021}, making them interesting objects for our comparison. \\
One caveat to this comparison is that asteroids usually have very few to no particles in suspension around their surface, in contrast to cometary dust and what we would expect for \HR. As discussed in Sect.\,\ref{sec_lab_mes}, this can have an impact on the polarimetric properties of the particles; however, in this work, we find that this effect is lessened due to the high absorbance of the particles.

\begin{figure}[ht!]
\centering
\includegraphics[width=\hsize]{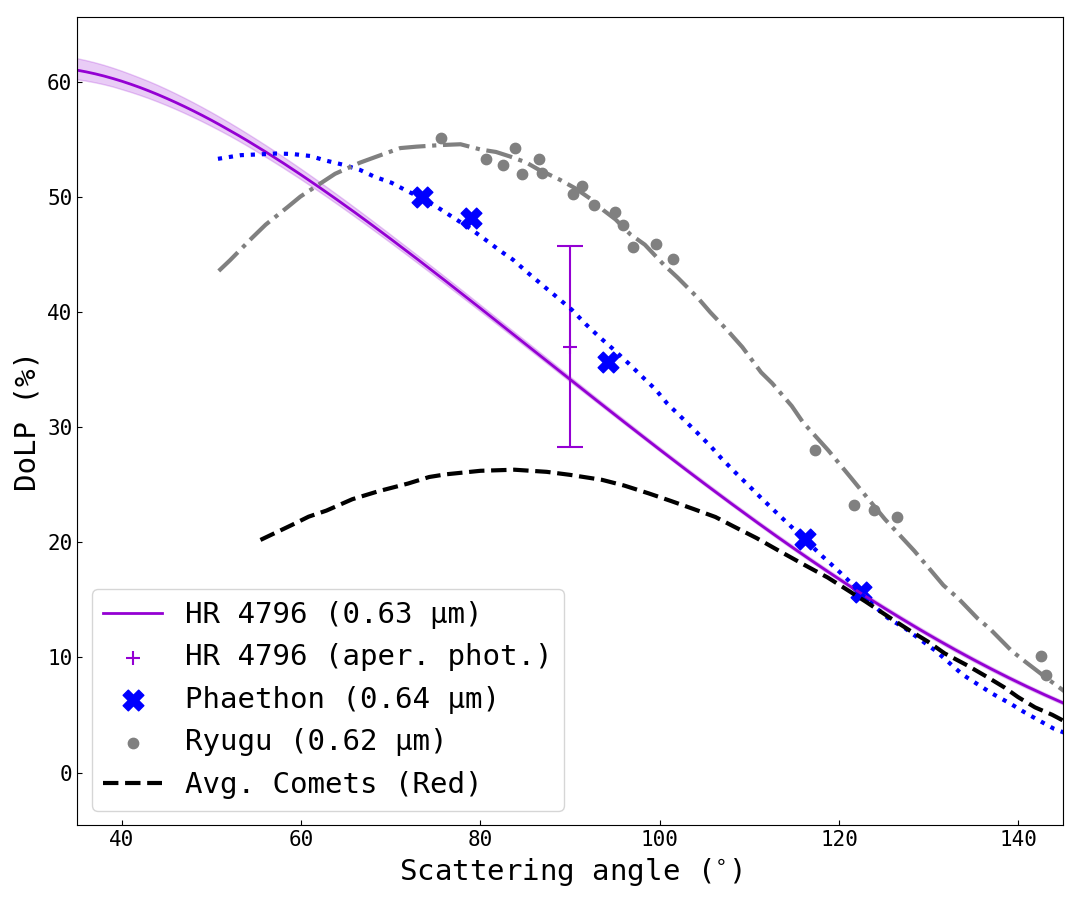}
    \caption{Comparison of the obtained DoLP for \HR{} at $\lambda$ = 0.63 µm, both parametric (plain violet line) and the photometric value at 90° (violet cross) with the asteroids Phaethon (B-type, blue dotted line and crosses); Ryugu (C-type, grey dashed line and points); and an average value for Solar System comets in the red domain (black dotted line). The data for Ryugu are from \citet{Kuroda_2021,Hadamcik_2023}, the points of Phaethon are from \citet{Ito_2018}, while the curve is from \citet{Hadamcik_2023}. The average comet value in the red domain ($\sim0.65$ µm) is from \citet{Kiselev_2015}.}
    \label{fig_HR_SS}
\end{figure}

\begin{table}[ht!]
\renewcommand{\arraystretch}{2} 
\caption{Maximum value, $P_M$, and scattering angle, $\theta_M$, of the DoLP peak for \HR{} (this work) and comparison with Solar System objects.}               
\label{table_doLP_comp_SS}   
\centering                        
\begin{tabular}{c | c | c | c  }      
\hline        
Object   & $\lambda $ & $P_M$ & $\theta_M$ \\
\hline
\HR{}    & 0.63 µm & $\sim$61 \% & $\sim$30° \\
\hline
Comets\tablefootmark{a} (avg)    & Red ($\sim$0.65 µm)& $\sim$26 \% & $\sim$83° \\
\hline
Ryugu\tablefootmark{b}    & 0.62 µm & $\sim$55 \% & $\sim$70° \\
\hline
Phaethon\tablefootmark{c}  & 0.64 µm & $\geq$ 50 \% & $\sim$60° \\
\hline
\HR{}    & 1.25 µm & $\sim$47 \% & $\sim$50° \\
\hline
Comets\tablefootmark{a} (avg) & NIR (1.25-2.2 µm) & $\sim$25 \% & $\sim$86° \\
\hline                                  
\end{tabular}
\tablefoot{
\tablefoottext{a}{Average comets data in the visible and NIR range from \citet{Kiselev_2015}} 
\tablefoottext{b}{Ryugu data from \citet{Kuroda_2021} and \citet{Hadamcik_2023}.} 
\tablefoottext{c}{Phaethon data from \citet{Ito_2018} and \citet{Hadamcik_2023}.}}
\end{table}

In Fig.\,\ref{fig_HR_SS} and Table \ref{table_doLP_comp_SS}, we compare the DoLPs of \HR{} at several wavelengths to the DoLPs of Phaethon, a B-type asteroid, Ryugu, a C-type asteroid, and an averaged value for the DoLP of Solar System comets in the visible ($\sim$0.65 µm) and NIR (1.25 µm, 1.65 µm, 2.2 µm). Both in the visible (Fig.\,\ref{fig_HR_SS}) and NIR (Table \ref{table_doLP_comp_SS}), the DoLPs of \HR{} and the averaged value for comets are not compatible; whereas in the visible, more processed asteroids such as Phaethon or Ryugu peak at values closer to \HR. It is also interesting to note that the nonparametric value of \HR's DoLP that we obtained at 90° is incompatible with the comet value at the same angle, but compatible within the error bars with Phaethon's DoLP at 90°. \\
Using the data in Fig. \ref{fig_HR_SS} and Table \ref{table_doLP_comp_SS}, we were able to conclude that \HR's DoLP is closer to objects that have gone through more altering processes than comets. Although $\theta_M$[asteroids] remains bigger than $\theta_M$[\HR], the general trend agrees with a composition similar to that of low albedo asteroids for \HR{} rather than a cometary composition. This would tend to indicate that \HR{} is a dynamically cold disk, with already processed and devolatized material (contrary to HD~181327 \citealp{Xie_2025}). This hypothesis of a cold disk consisting of highly processed material is consistent with the findings and discussions in \citet{Lisse_2017} when performing IR spectroscopy on \HR. \\
Nevertheless, it is important to emphasize that although the DoLP of \HR{} seems close to that of low albedo asteroids, this resemblance can not necessarily be extended to every optical properties. As shown on Fig.\,\ref{DoLP_90_All}, we obtained a blue (negative) polarimetric spectral slope in \HR{} in the visible and NIR range, whereas objects consisting of more porous aggregates, such as cometary dust usually show a red or grey polarimetric spectral slope \citep{Kolokolova_2004, Bagnulo_2024}, adding value to our hypothesis of a processed, devolatized disk. However, although a blue polarimetric spectral slope can be observed on some C-type asteroids \citep{Kwon_2023}, it is rarely to the extent of \HR{} \citep{Bagnulo_2015} and not a general rule; for example, Phaethon shows a red polarimetric slope \citep{Kiselev_2022}. \\
Additionally, when considering the spectral reflectance, we obtained a red slope in \HR{} (see Fig.\,\ref{Refl_90_All}) that resembles the one obtained on Ryugu \citep{Kitazato_2019}, but not necessarily on Phaethon. In particular, Phaethon presents a variable spectral slope, sometimes blue \citep{Kareta_2018} and sometimes red or grey (\citealp{Kiselev_2022} and references therein). This highlights the need for the comparison of multiple optical properties, at numerous wavelengths and scattering angles, to constrain the properties of the dust. \\
As discussed in Sect.\,\ref{subsec_DoLP_Res} and Sect.\,\ref{subsec_Refl_Res}, our results tend to indicate large particles, of a few micrometers in size to s < 100 µm as the main scatterers over the observation range. However, the opaque minerals particles that are responsible for the darkness of carbonaceous chondrites are usually subµm inclusions in the fine grain matrix of these rocks. Therefore, \HR's dust could be made of such large metallic particles or its properties could result from space weathering processes (solar wind irradiation, micro-meteoritic impacts, etc.) melting and devolatizing the surfacial layer of carbonaceous chondrite-like matrix aggregates, as observed on samples returned from Ryugu by Hayabusa 2 \citep{Noguchi_2023,Hiroi_2023}.

\subsection{Summary}
\label{sec_ccsion}
We extracted the SPFs and DoLPs of \HR{} at three new wavelengths using SPHERE, both in the optical and NIR range. Using a novel parametric joint foward modeling approach, we obtained a geometric description of the disk that is roughly consistent with previous findings, as well as a tentative constraint on the opening angle of the disk. Our results also allow to confirm the high degree of linear polarization at small scattering angles observed by \citet{Arriaga_2020} for $\lambda\sim$ 2.05 µm. \\
We performed aperture photometry in the ansae of the disk ($\theta$ = 90°) and found a negative polarimetric spectral slope. We also observed a red slope in the spectral reflectance of \HR{}, with values compatible to the ones obtained previously with a range of instruments \citep{Milli_2017, Rodigas_2015, Kueny_2026}. \\
We compared our results to experimental measurements of the DoLP and the spectral reflectance of iron sulfide particles of different sizes. We find that the DoLP and spectral slope of \HR{} we obtained are both compatible with compact opaque minerals of a size ranging from a few micrometers to 100 µm being the main scatterers in the disk, matching the conclusions from \citet{Kueny_2026} where they find that large (several µm) absorbing particles dominate the SPF. However, as modeled in \citet{Rodigas_2015}, or \citet{Chen_2020}, the precise composition of the disk is very likely a mixture of various elements (astrosilicates, carbonaceous compounds, etc.), and very challenging to disentangle. Further laboratory experiments, modeling, and observations (e.g., at longer wavelengths to observe solid-state features) are needed to help constrain the dust properties even further.\\
Finally, we compared the high $P_M$ and small $\theta_M$ DoLP of \HR{} to those of Solar System objects and found that the polarimetric properties of the disk are closer to those of highly processed low-albedo asteroids than to those of comets. The negative slope of the DoLP at 90° as a function of the wavelength is also compatible with compact particles, whereas porous aggregates as seen in cometary dust usually show an increase of the DoLP with the wavelength \citep{Hadamcik_2003, Kolokolova_2004}. We also discuss on space weathering processes, similar to those that occurred on the surface of Ryugu, to explain the possible presence of large opaque minerals particles at the surface of the disk's dust. 

\section*{Data availability}
The VLT/SPHERE data products presented in Fig\ref{fig_Obs_all} are available in electronic form as FITS files at the CDS via anonymous ftp to cdsarc.u-strasbg.fr (130.79.128.5) or via \url{http://cdsweb.u-strasbg.fr/cgi-bin/qcat?J/A+A/}. \\
The laboratory measurements made using the \texttt{SHADOWS} instrument can be found on the \href{https://www.sshade.eu/db/ghosst}{\texttt{GhoSST}} database, hosted on the \href{https://www.sshade.eu}{\texttt{SSHADE}} database infrastructure: \\ 
-DoLP\textsubscript{Q} measurements: 
\url{https://doi.org/10.26302/SSHADE/EXPERIMENT_LB_20260917_001} \\
-spectral reflectance measurements: \url{https://doi.org/10.26302/SSHADE/EXPERIMENT_LB_20260909_001}.

\begin{acknowledgements}
     M.B. acknowledges funding from the Agence Nationale de la Recherche through the DDISK project (grant No. ANR-21-CE31-0015, P.I: M. Langlois). M.B., J.M. and O.P. acknowledge funding from the PNP (French National Planetology Program) through the EPOPEE project. We acknowledge Louis Wieczorek, Lucas Patty and Stefano Spadaccia for the measurements shown in Fig. C.5. O.P., M.B., J.M. acknowledge financial support from the Centre national d’études spatiales (CNES), France (ROR: https://ror.org/04h1h0y33) within the framework of the Comet Interceptor mission. This work has made use of the High Contrast Data Centre, jointly operated by OSUG/IPAG (Grenoble), PYTHEAS/LAM/CeSAM (Marseille), OCA/Lagrange (Nice), Observatoire de Paris/LESIA (Paris), and Observatoire de Lyon/CRAL, and supported by a grant from Labex OSUG@2020 (Investissements d’avenir – ANR10 LABX56). G.D., L.M., M.R. and J.M. acknowledge funding from the European Research Council (ERC) under the European Union’s Horizon Europe research and innovation program (grant agreement No. 101053020, project Dust2Planets, PI: F. Menard). J.M. thanks the Swiss National Science Foundation for financial support under grant number P500PT 222298. RT was supported by JSPS KAKENHI grant Number JP25K07351.
\end{acknowledgements}

%\begin{thebibliography}
\bibliographystyle{bibtex/aa}
\bibliography{bibtex/Bib_Sauron}
%\end{thebibliography}

\begin{appendix}

\section{Model}
\label{sec_caveats}
\subsection{Disk morphology}
\label{app_disk_morpho}
Following an earlier approach to modeling \HR{} \citep{Milli_2017, Milli_2019}, we used a simple geometric model (isotropic) assuming an elliptical disk, with the reference radius, $r_0$, expressed in polar coordinates as
\begin{equation}  
r_0(\varphi) = \frac{a\;(1-e^2)}{1 - e\cos (\varphi - \omega)},
\label{eqR_ellipse}
\end{equation}
with $e$ as the ellipticity of the disk, $\omega$ as the argument of the pericenter, and $a$ as its semi-major axis (in arcsec).
Following \citet{Augereau_1999}, the dust volume density distribution was parametrized using a smoothly connected double power law, with
\begin{equation}  
\rho(r,\varphi,z) = \rho_0\ \left( \frac{2}{\left( \frac{r}{r_0(\varphi)}\right)^{-2 \; \alpha_{in}} + \left( \frac{r}{r_0(\varphi)}\right)^{-2\:\alpha_{out}} }\right)^{1/2} \; e^{-\left(\frac{z}{H(r)}\right) ^{\gamma}},
\label{eq_dens_distr}
\end{equation}
in cylindrical coordinates. The parameter $\alpha_{in}$ and $\alpha_{out}$ are respectively the inner and outer slopes of the dust volume density distribution, $\gamma$ describes the shape of the vertical distribution, $\rho_0$ the density at the reference radius $r_0$ and in the midplane ($z =0$), and $H(r)$ is the scale height of the disk, defined as
\begin{equation}  
H(r,\varphi) = \xi_0 \;  \frac{r}{r_0(\varphi) }
\label{eq_sca_height},
\end{equation}
with $\xi_0$ the scale height at the reference radius $r_0$, and $\beta$ the flaring index of the disk. The disk vertical opening angle $\Psi$ is also defined as $\Psi = \xi_0 / a$.\newline
Following prior modeling of the disk \citep{Milli_2015,Milli_2017,ARKS_III}, the vertical profile of the disk is assumed to be Gaussian ($\gamma = 2$), while the flaring of the disk is considered linear ($\beta =1$).

\subsection{Value of $\alpha_{in}$}
\label{app_alphain}
\begin{figure}[ht!]
\centering
\includegraphics[width=\hsize]{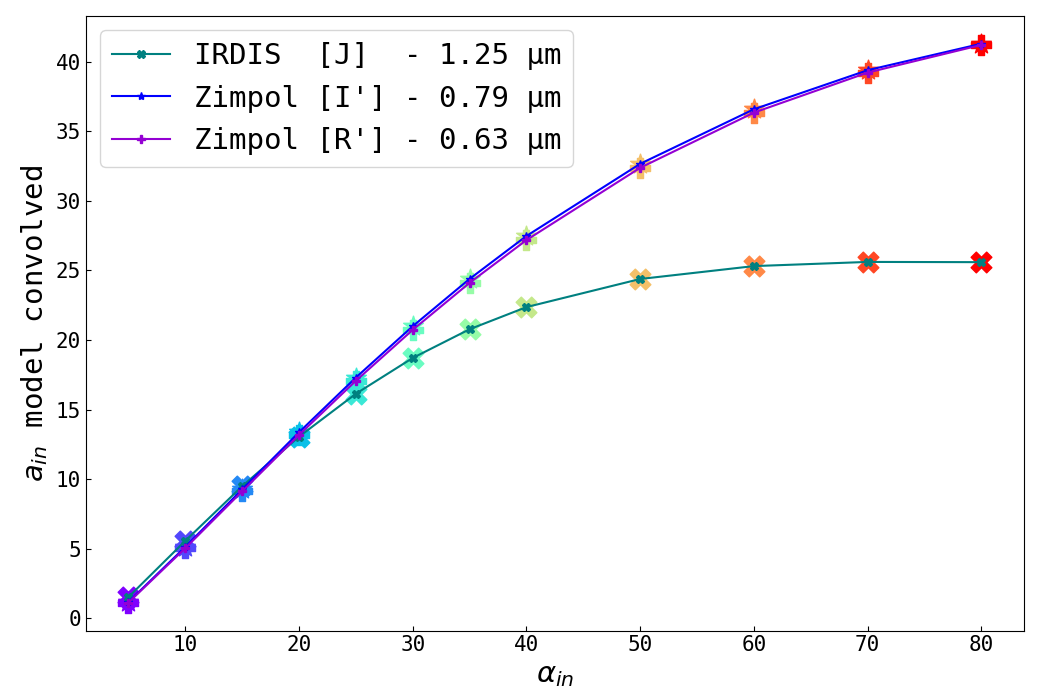}
    \caption{Measured slope, $a_{in}$, on the convolved model as a function of the initial $\alpha_{in}$ value of the model, for ZIMPOL [R'] (violet), ZIMPOL [I'] (blue) and IRDIS [J] (teal).}
    \label{fig_alphain_ainconv}
\end{figure}

The inner slope of \HR{} is notoriously steep (see $\alpha_{in}\sim 34.5$ in \citealp{ARKS_V}, $\alpha_{in}\sim 23.3$ in \citealp{Milli_2017}, $\alpha_{in}\sim45$ in \citealp{Kueny_2026}), and difficult to constrain. When left as a free parameter using the IRDIS data, initial modeling showed that $\alpha_{in}$'s value tended to increase: when the model was stopped we had $\alpha_{in}\sim 42$ (for an initial value of 18).  \\
Due to the convolution by the PSF of the instrument, high values of $\alpha_{in}$ cannot be distinguished from one another. To determine this threshold value, we generated several disks, mimicking \HR's parameters, albeit with varying $\alpha_{in}$ values (5; 10; 15; 20; 25; 30; 35; 40; 50; 60; 70; 80). Each model was then convolved with the PSF corresponding to each of the three data set. As the inner slope measured on the profiles of these models, $a_{in}$, is linked to the value of $\alpha_{in}$, we compared these values. Fig. \ref{fig_alphain_ainconv} shows the slope ($a_{in}$) measured on the convolved models as a function of the value of $\alpha_{in}$ of the model. This figure shows that for the IRDIS data, we cannot distinguish between models with $\alpha_{in}$ values above $\sim 45$ due to convolution effect, as the measured slope plateaus. This effect happens for higher values of $\alpha_{in}$ in the case of ZIMPOL, not plotted on the graph, though we can see the beginning of the same effect taking place. Additionally, these plateau values are minima, as we only added the convolution effect, and not the noise or derotation of the images, which would tend to further diminish the $\alpha_{in}$ plateau value. 

Due to the difficulty in constraining $\alpha_{in}$, we decided to fix it in our models. To determine the value, we measured the slopes $a_{in}$ of the profiles of the $Q_{\phi}$ data for ZIMPOL Rp, ZIMPOL Ip, and IRDIS. We obtained different values for each ansa (as expected, see \citealp{Milli_2019}), with the northern ansa's slope being steeper than the southern one. We averaged over the two ansae then reported those values in Fig. \ref{fig_alphain_ainconv}, to recover the corresponding $\alpha_{in}$. We obtained $\alpha_{in}$(ZIMPOL[R']) $\sim$ 34, $\alpha_{in}$(ZIMPOL[I']) $\sim$ 30, and $\alpha_{in}$(IRDIS[J]) $\sim$ 32. For simplification purposes, we decided to use $\alpha_{in} = 30$ as a fixed parameter in all our models. 

\subsection{Value of $\Psi$}
\label{app_opang}
When running the total intensity models on the IRDIS data, it became apparent that the model could not constrain properly the opening angle, $\Psi$. This is most likely due to the extreme thinness of the disk combined with the self-subtraction in the disk induced by the ADI \citep{Milli2012}, visible close to the star (see the upper third panel in Fig. \ref{fig_Obs_all}).
This led to the total intensity model tending towards an extremely small value of $\Psi$ ($\sim$ 0.3 \%), although on nonconverged models. This is a similar problem to that encountered in other works when trying to constrain \HR's opening angle: the best-fit model of \citet{Chen_2020} had a scale height of < 0.001 and \citet{Kueny_2026} had to fix their opening angle at 1\% for their RDI-KLIP images. However, when we ran the model on the polarized IRDIS data, where there is no self-subtraction, we managed to obtain an opening angle that converged to $\Psi \sim 1.22\%$. We decided to fix $\Psi \geq 1.20\%$ for the IRDIS joint modeling, and obtained $\Psi = 1.20\% ^{+0.01}_{-0.00}$. To diminish the number of parameters for the ZIMPOL modeling (already time consuming), we then fixed  $\Psi = 1.20\%$ for both ZIMPOL R' and I'.

\subsection{Parametric description of the phase functions}
\label{param_descr_phf}
To assess the validity of our parametric description, we used another, nonparametric description of our model. In this description, the SPF and pSPF were recovered by adding the (polarized) intensity of the disk at chosen scattering angles as varying parameters in the model. Following Fig. 12 in \citet{Perrin_2015}, we estimated that the accessible scattering angles in \HR{} were in the range of [13° to 167°]. To sample the value of the phase function, we chose the following angles: [13°, 28°, 54°, 90°, 136°, 155°, 167°] for IRDIS in total and polarized intensity. For the joint models, this amounts to 21 free parameters (7 morphological parameters presented in \ref{model_morpho}, flux values at seven angles for the total intensity images, and identically for the polarized intensity). Due to the high number of free parameters, we performed simple Nelder-Mead minimization, using the \href{https://docs.scipy.org/doc/scipy-1.17.0/reference/optimize.html}{\texttt{scipy.optimize}} package for the IRDIS data to obtain these values at the chosen scattering angles. We then fit these discrete data points (still using the minimization function) a double Henyey-Greenstein on the total intensity anchor points and a beta function on the DoLP anchor points (corresponding to the polarized intensity points over the total intensity points).

\begin{figure}[ht!]
\centering
\includegraphics[width=\hsize]{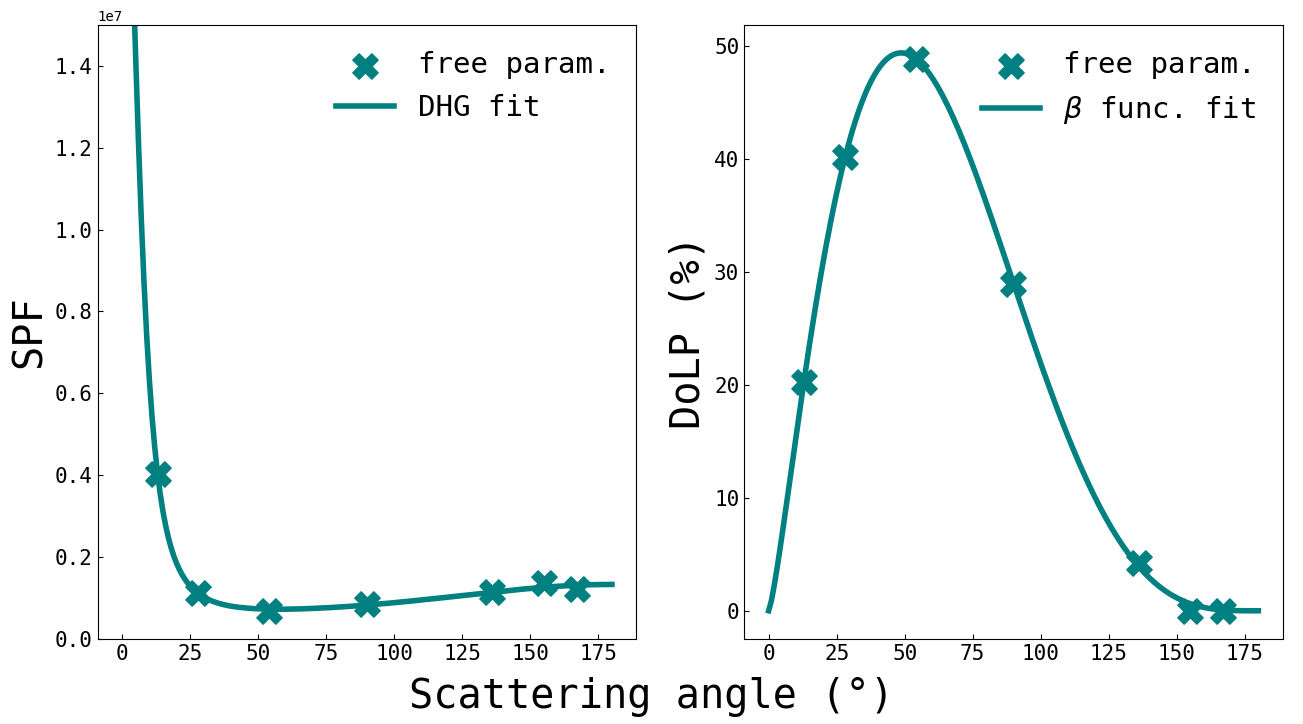}
    \caption{IRDIS data. Scattering phase function (left) and degree of linear polarization (right) obtained by selecting specific angles and using the corresponding intensities as free parameters in the model (crosses) and curves obtained by fitting a double Henyey-Greenstein (left) and a beta function (right) to those values.}
    \label{fig_param_obs}
\end{figure}

In Fig.\ref{fig_param_obs}, we can see that for the IRDIS data, the two-component Henyey-Greenstein and the beta function seem to be a good fit for the SPF and DoLP (respectively) that we obtained using the nonparametric description. Building on those results, we used this parametric description for our modeling of the phase functions of the disk, as it allows to reduce the number of free parameters from 21 (seven morphological, seven anchor points for the SPF, and seven anchor points for the pSPF) to 14 (seven morphological, four for the SPF, and three for the DoLP).

\section{Results}

\subsection{DoLP - beta function}
\label{app_param_beta_func}
The best fit-parameters for the beta function used to parametrize the degree of linear polarization given in Fig.\,\ref{fig_dolp_obs} are presented in Table \ref{table_pspf_param}. This description was used following the parametrization in \citet{Ren_2023}, where they applied it for numerous protoplanetary disks. 

\begin{table}
\renewcommand{\arraystretch}{2} 
\caption{Beta function parameters obtained for the ZIMPOL and IRDIS models.}           
\label{table_pspf_param}   
\centering                        
\begin{tabular}{c | c | c | c }      
\hline               
Parameter & ZIMPOL (R') & ZIMPOL (I') & IRDIS (J)  \\         
\hline
$\lambda$ & 0.63 µm & 0.79 µm & 1.25 µm \\         
\hline 
$\alpha$  & $1.42^{+0.03}_{-0.03}$ & $1.78^{+0.07}_{-0.08}$ & $2.40^{+0.04}_{-0.04}$   \\
$\beta$   & $3.04^{+0.04}_{-0.04}$ & $3.50^{+0.08}_{-0.09}$ & $4.45^{+0.06}_{-0.06}$   \\    
$f_M$     & $0.61^{+0.01}_{-0.01}$ & $0.57^{+0.01}_{-0.01}$ & $0.47^{+0.01}_{-0.01}$   \\
\hline                                  
\end{tabular}
\tablefoot{The error bars are shown at 3$\sigma$.}
\end{table}

\begin{figure}[ht!]
\centering
\includegraphics[width=0.96\hsize]{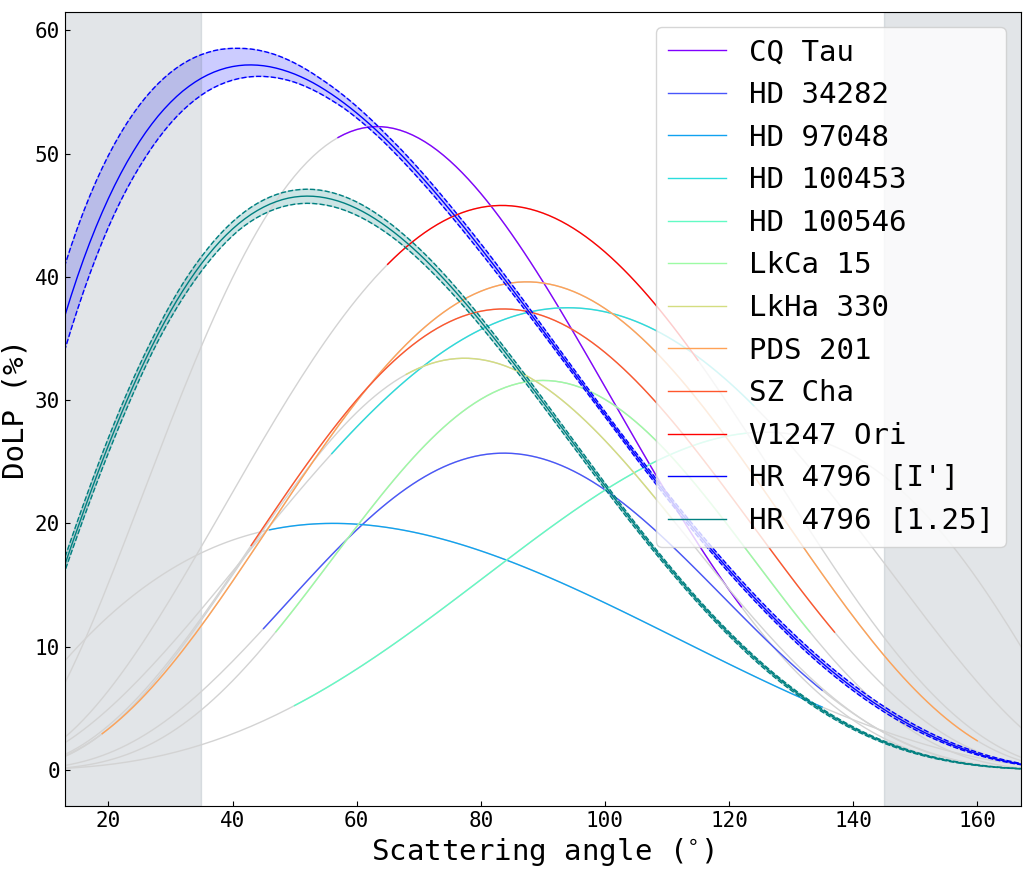}
    \caption{Parametrization of the DoLP using the beta function for \HR{} (at $\lambda$ = 0.79 µm in dark blue and $\lambda$ = 1.25 µm in teal). The DoLPs obtained by \citet{Ren_2023} for numerous protoplanetary disks are overplotted (in various colours). The shaded area corresponds to the scattering angles accessible in the IRDIS data, but not the ZIMPOL ones, and the shaded curves are the scattering angles not accessible for each protoplanetary disks.}
    \label{fig_HR_PPD}
\end{figure}

In Fig.\,\ref{fig_HR_PPD}, we plot the DoLPs we obtained at 0.79 µm and 1.25 µm, and added the DoLPs from the protoplanetary disks in \citet{Ren_2023} in various colours. It is interesting to note that \HR's polarimetric properties are different from those of most protoplanetary disks: in \citet{Ren_2023}, they only found one disk with $f_M$>47\% (CQ Tau) and no disk with a $\theta_M$ < 50°. The smallest $\theta_M$ they reported is 56.3° for HD~97048. 

\subsection{Aperture photometry}
\label{app_aper_phot}

We used the \href{https://photutils.readthedocs.io/en/2.3.0/index.html}{\texttt{PHOTUTILS}} package \citep{Bradley_2025} to select circular apertures at $\sim$ 1 arcsec of the central star, centered in each ansa, with a 63 mas radius as chosen in \citet{Milli_2017}, also corresponding to 1.5 resolution element of GPI H. We calculated the photometry in these apertures (in each ansa, for the total intensity images and polarized light). We corrected these values by estimating the background subtraction as well as the throughput. \\
To estimate the error bars on the photometry in each ansa, we performed aperture photometry on nonoverlapping apertures of the same radius, distributed on a circle with a radius equal to the semi-major axis of the disk, with the exception of the ansae. We then calculated the standard deviation of the obtained apertures (for both total and polarized intensity). We then estimated the error on the background subtraction through a similar process and obtained our uncertainties using the propagation of the uncertainties. \\
For the spectral reflectance, we divided the photometry in the ansae in total intensity by the stellar flux, estimated using the noncoronagraphic image of the star (PSF). We then divided this result by the surface of the aperture to obtain the result in contrast per arcsec squared. Finally, the error bars are estimated as previously for the photometry of the disk's ansae, and combined with an estimated 10\% error on the total star flux. 

\section{Laboratory measurements}

\subsection{Characteristics of the iron sulfide sample}
\label{app_FeS_MEB}
\begin{figure}[ht!]
\centering
\includegraphics[width=0.9\hsize]{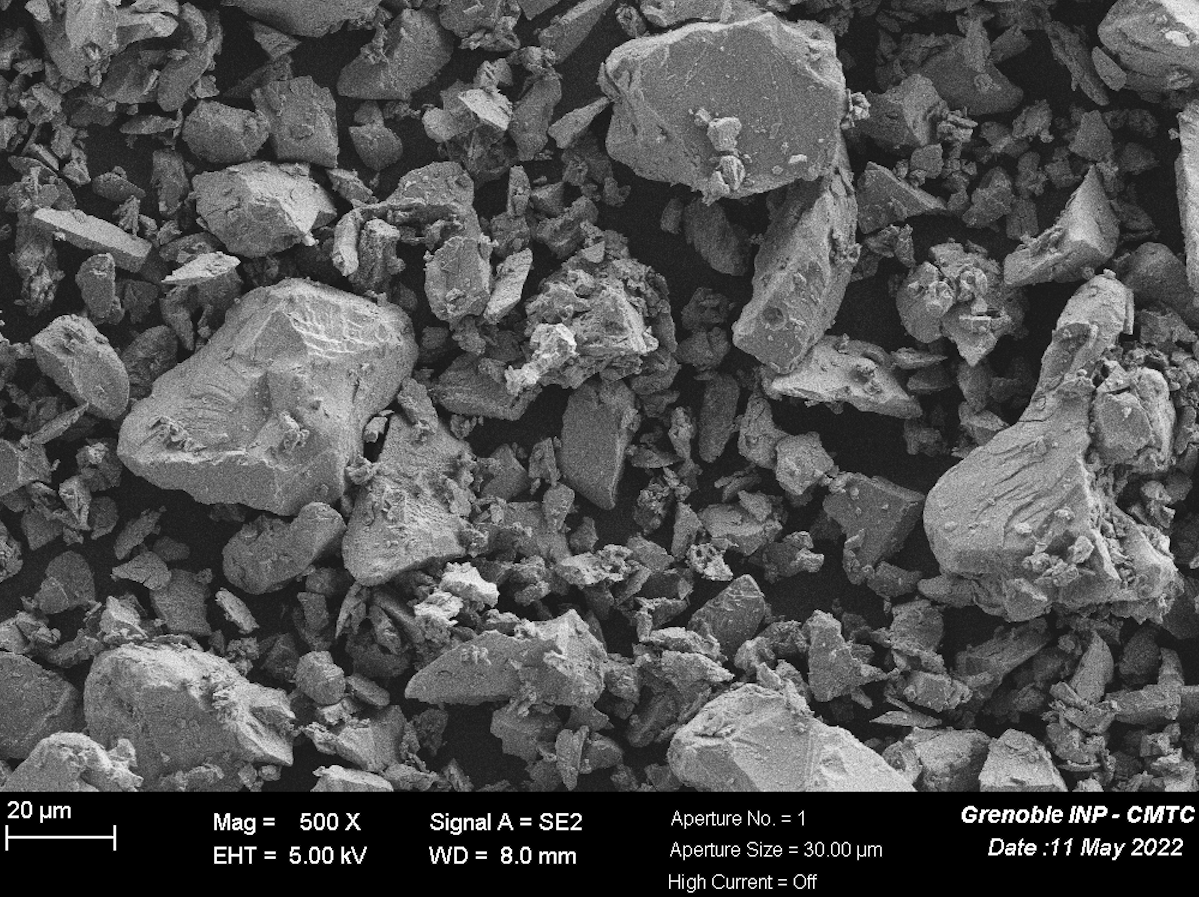}
    \caption{Scanning Electron Microscope (SEM) image of the iron sulfide sample with large (FeS s < 100 µm) used in our laboratory measurements.}
    \label{Fig_FeS_MEB}
\includegraphics[width=\hsize]{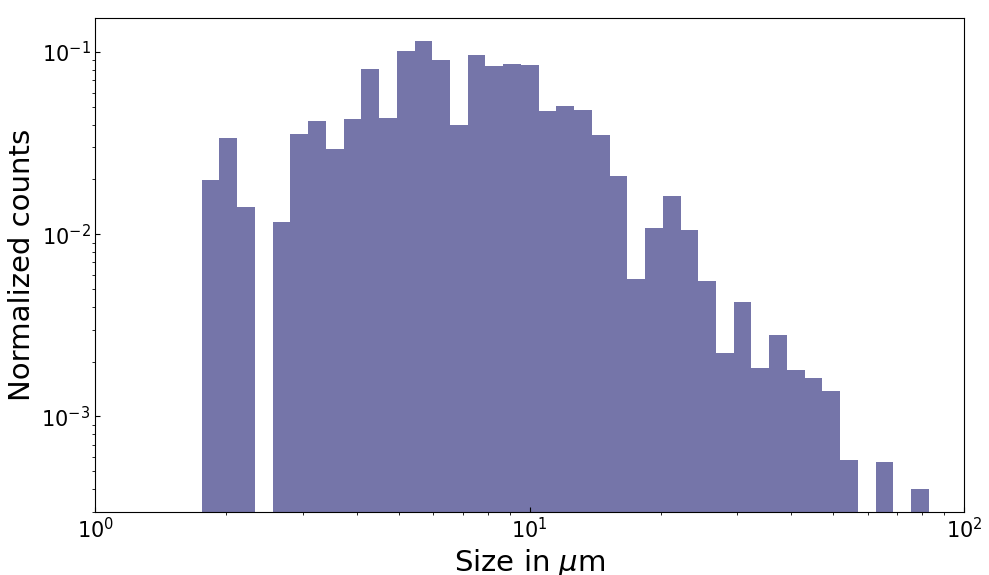}
    \caption{Size distribution of the particles of the large particles (s < 100 µm) FeS sample}
    \label{fig_size_distr_FeS}
\end{figure}

The iron sulfide sample we worked on was purchased from Alfa Aesar (ref. A15569.0B). A X-ray diffraction measurement performed on this commercial powder indicates it is composed of of 55 vol\% of troilite FeS and at 45 vol\% of pyrrhotite Fe\textsubscript{1-x}S\textsubscript{x} (with 0 < x < 0.2), with particles ranging from a micrometers in size to 100 µm. The characteristics of this FeS (s < 100 µm) sample are shown on Fig.\,\ref{Fig_FeS_MEB}: Fig.\,\ref{fig_size_distr_FeS}

\begin{figure}[ht!]
\centering
\includegraphics[width=0.9\hsize]{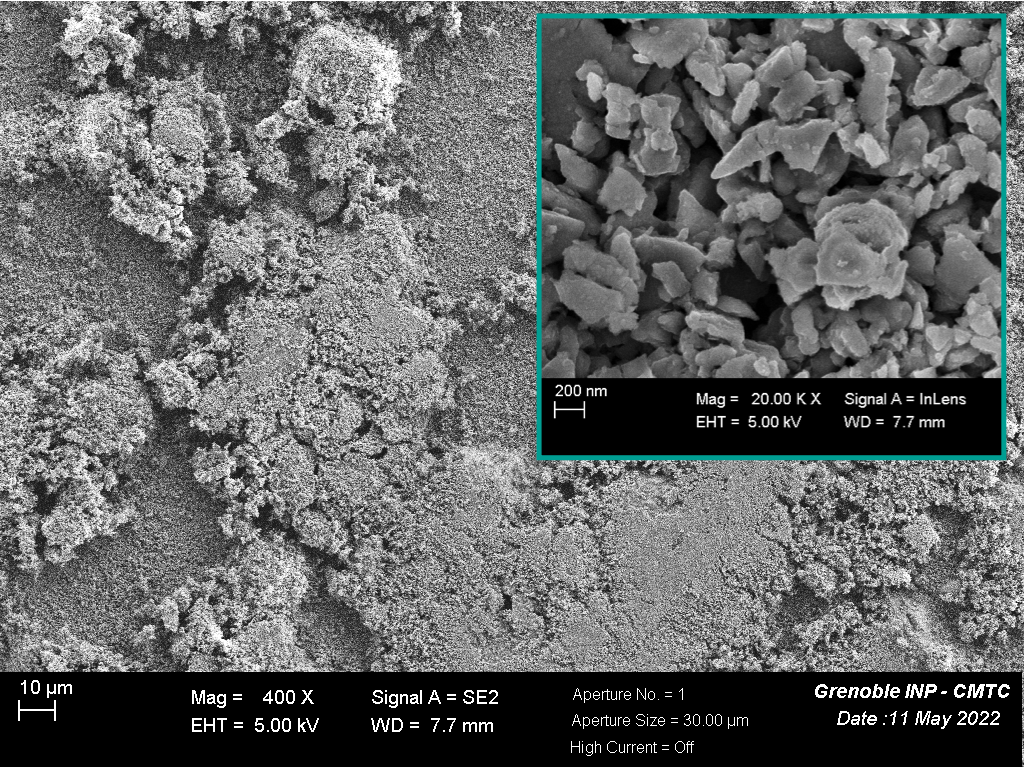}
    \caption{Scanning Electron Microscope (SEM) image of the iron sulfide sample with small particles (FeS s < 1 µm) used in our laboratory measurements. The upper right inset shows the same sample observed at a smaller scale (200 nm against 10 µm).}
    \label{Fig_FeS_MEB_small}
\end{figure}

The particles are relatively big, ranging from a few micrometers to $\sim$ 90 µm. The size distribution peaks between 5 and 10 µm. The SEM image shows that the particles are compact, and have rough edges (rather than smooth). \\
This sample was also ground down to a small powder of submicronic particles. The characteristics of this sample are described in \citet{Sultana_2023} (Table 1 and supplementary figure 3), and SEM images of this second sample (FeS s < 1 µm) are shown in Fig.\,\ref{Fig_FeS_MEB_small}. This figure has a similar magnification than shown on Fig.\,\ref{fig_size_distr_FeS}, and shows that this second sample has much smaller particles (mean particle size $\sim$ 0.3 µm \citealp{Sultana_2023}). The upper right inset shows an even higher magnification, displaying the fine structure of the particles.

\subsection{Experimental setup}
\label{exp_setup}
\href{https://cold-spectro.sshade.eu/-SHADOWS-Micro-Spectro-Gonio-Radiometer-}{\texttt{SHADOWS}}, built at the IPAG laboratory, measures bidirectional reflectance spectra, from 0.50 to 4.8 µm, of dark and small samples such as meteorites or natural/synthetic mineral samples with analogous properties to asteroidal surfaces \citep{Sultana_2023}. The principle is as follows: monochromatic light is produced from an halogen lamp and a monochromator before being injected into a bundle of fibres. The spectro-gonio-radiometer consists of two arms: one (illumination arm) holding the optical fibres and a mirror sending monochromatic light onto the sample, and another arm (observation arm) holding two detectors (visible and IR range) that collects the light scattered by the sample. Both the illumination and observation arms are motorized, allowing to measure reflected light in a wide range of scattering angles (from 20 to 175°). The star light illuminating circumstellar dust is unpolarized. To measure reflectance spectra and polarimetric phase curves of dust samples under similar conditions with SHADOWS, we need to depolarize the light before it illuminates the sample (as the lamp, the monochromator, the mirrors, etc., can induce polarization). A first depolarizer is placed before the light is injected into the optical fibers (Thorlabs DPP, LCP Achromatic Depolarizer, Ø1", uncoated) and a second depolarizer is placed on the optical path after the reflection of the light from the fiber on the mirror of the illuminance arm (Thorlabs DPU-25, Quartz-Wedge Achromatic Depolarizer). These optics spatially vary the polarization of the beam (on a scale of 25 µm for DPP, and 2 mm for DPU), which produces a pseudo-random polarization. To spatially average the effect of the incident beam on the surface of the sample, the DPU polarizer is rotated at 2 Hz, twice as high as the time it takes for the detectors to integrate the signal. Finally, a polarizer was added in front of each of the detectors, so as to measure the degree of linear polarization of the sample. \newline 
The configuration that was chosen for our measurements was specular, namely, the incident and emergence angles are equal for each of the measured configurations. The range of both arms allowed us to measure the scattering angles varying from 20° to 140° for each wavelength, with a step of 10° between 20° and 60°, and a step of 20° between 60° and 140°. The wavelength range was from 0.55 µm to 2.95 µm, with a step of 0.1 µm. The measurements were made with the analyzing polarizer placed at an angle of 0° then at an angle of 90°, to measure the Q Stokes parameter, defined as $Q = I_{90}-I_{0}$, with $I_{90}$ the intensity measured with the polarizer at a 90° angle (perpendicular to the scattering plane), and $I_{0}$ the one measured at 0° (parallel to the scattering plane). From these, we can obtain the degree of linear polarization of the sample, representing the proportion of light that is linearly polarized by the sample, and defined as
\begin{equation}
DoLP = \sqrt{\frac{Q^2 + U^2}{I^2}},
\label{eqDoLP}
\end{equation}
with $U$ the Stokes parameters such as $U = I_{135}-I_{45}$, and $I$ the total intensity. However, in our case, only Q was measured, as U is considered negligible. In this case, we can obtain the DoLP\textsubscript{Q}, with
\begin{equation}
DoLP_Q = \frac{Q}{I} = \frac{I_{90°}-I_{0°}}{I_{90°}+I_{0°}},
\label{eqDoLPQ}
\end{equation}

\subsection{Calibration and uncertainty sources}
\label{app_calib_shadows}

The calibration of the \texttt{SHADOWS} instrument (for total light intensity reflected from a sample surface, such as deposited particles) is presented in \citet{Potin_2018}. The overall error induced by the functioning of the instrument is estimated to be around 1\%, and was added in our calculation of the error bars of our measurements.\\
In comparison to the approach taken in this paper, we added a motorized stage that continuously rotates a quartz-wedge depolarizer (DPU-25) to ensure that the incident light is depolarized. When measuring the polarization of the incident light with the detectors of SHADOWS (by placing the illumination arm at 90° and each detector of the observation arm at 90°, see Fig.6 in \citet{Potin_2018}) we obtained the following DoLP:
\begin{figure}[ht!]
\centering
\includegraphics[width=0.96\hsize]{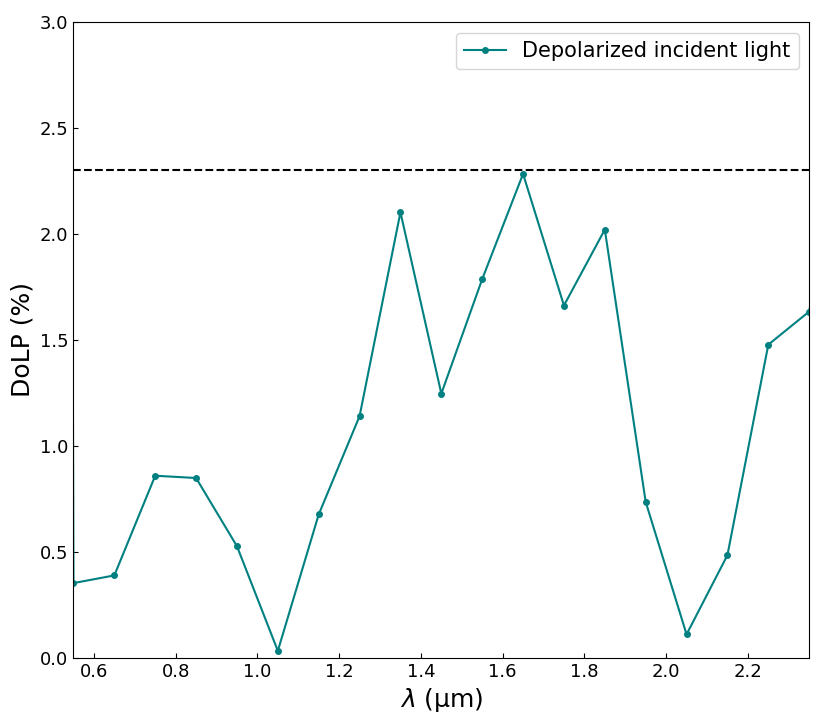}
    \caption{Measured residual degree of linear polarization of the incident light for $\lambda \in$ [0.55; 2.35] µm.}
    \label{Fig_Zero_Pol}
\end{figure}

Fig.\,\ref{Fig_Zero_Pol} shows that the residual polarization of the incident light is below 2.3\% over the whole wavelength range of interest in our study. This wavelength dependent residual incident polarization was also taken into account in our uncertainties  calculations.\\
We also added a polarizer before the detectors to measure the linear polarization of our samples. To ensure that our measurements are correct, we compared our results on the small (s < 1 µm) FeS sample at $\lambda$ = 0.64 µm using \texttt{SHADOWS} with the DoLP obtained on this sample at $\lambda$ = 0.625 µm using the \texttt{POLICES}\footnote{\url{https://www.space.unibe.ch/research/research_groups/planetary_imaging_group_pig/science/lossy/polices/index_eng.html}} experiment, hosted in Bern \citep{Poch_2018}.

\begin{figure}[ht!]
\centering
\includegraphics[width=0.96\hsize]{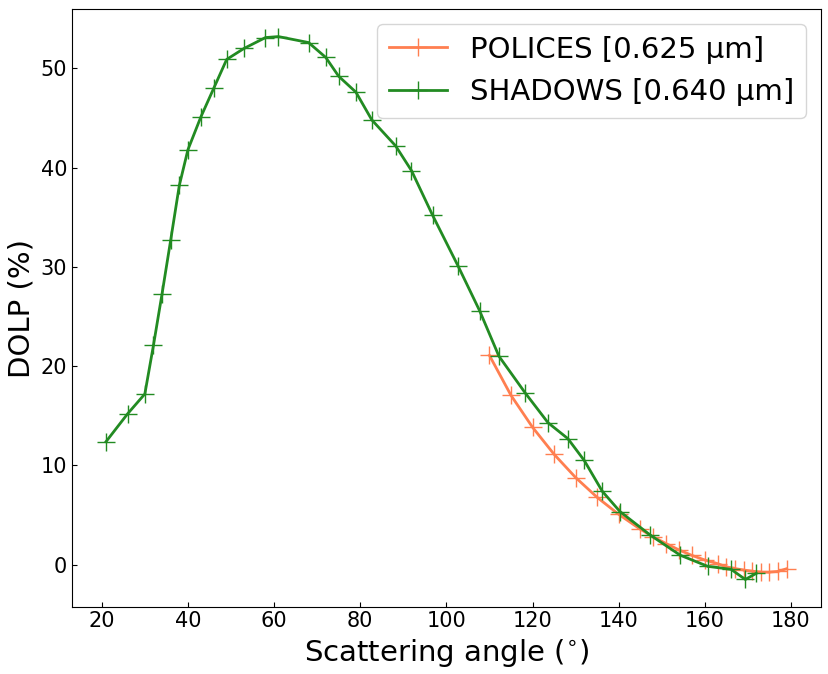}
    \caption{DoLP of an FeS sample with small particles (s < 1 µm) measured at 0.64 µm with \texttt{SHADOWS} and at 0.625 µm with the \texttt{POLICES} experimental setup.}
    \label{Fig_DoLP_comp_Shadows_polices}
\end{figure}

From Fig.\,\ref{Fig_DoLP_comp_Shadows_polices}, we consider that \texttt{SHADOWS} is well calibrated also for polarimetric measurements of deposited samples. \\
As previously stated, the overall error of the instrument, and the impact of the residual incident linear polarization have been taken into account in our error bar calculations. The final source of uncertainty we are considering in our error bar is statistical: for each of the configurations (wavelength + geometry), the measurement is repeated 40 times, allowing to obtain a statistical error on the S/N from the detection system.

\section{S/N map and S/N residual maps}
\label{app_res_best_models}

\begin{figure*}[ht!]
\centering
\includegraphics[width=0.96\textwidth]{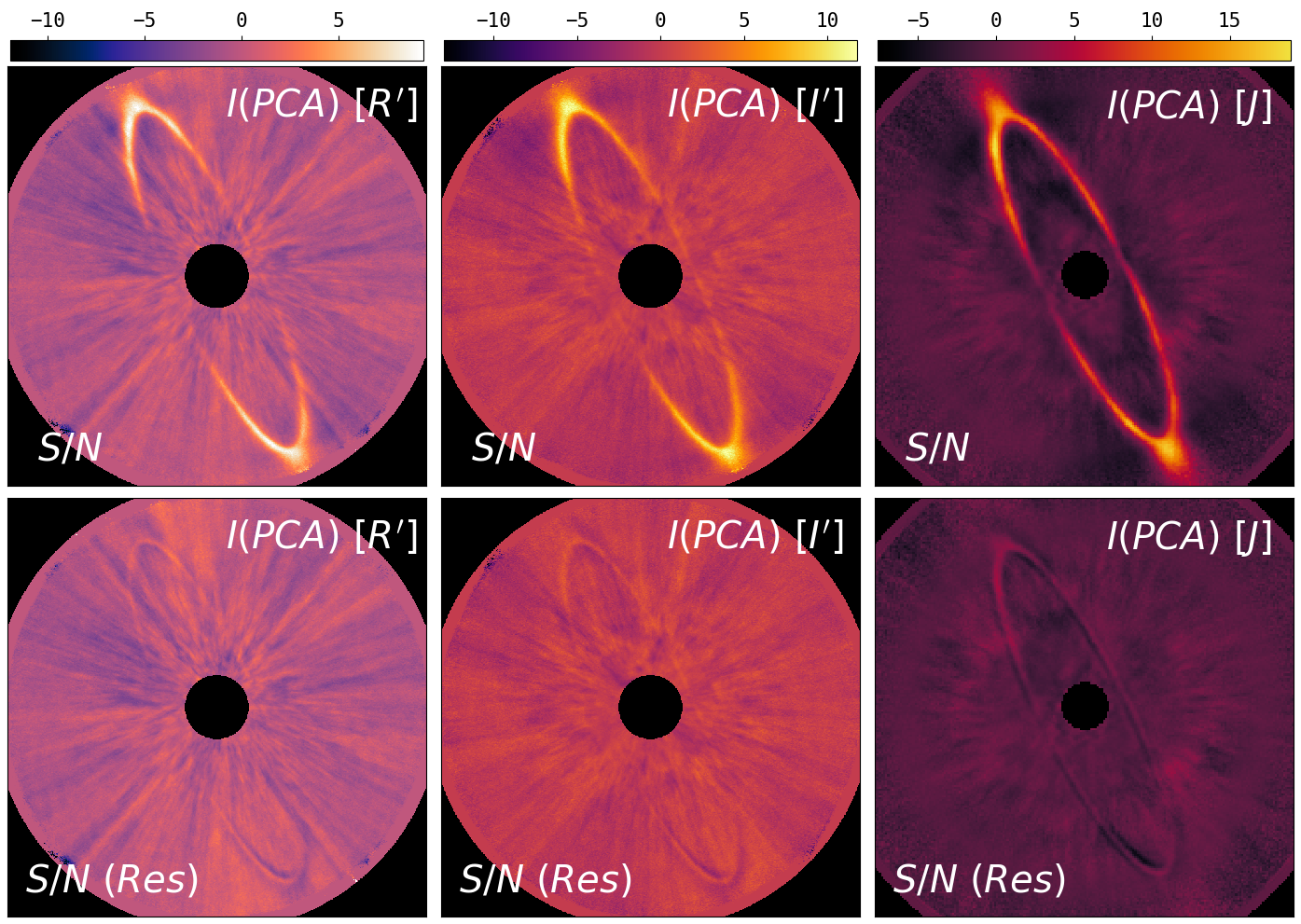}
    \caption{Total intensity: S/N map (top) and S/N of the residuals (bottom).}
    \label{fig_SNR_tot}
\includegraphics[width=0.96\textwidth]{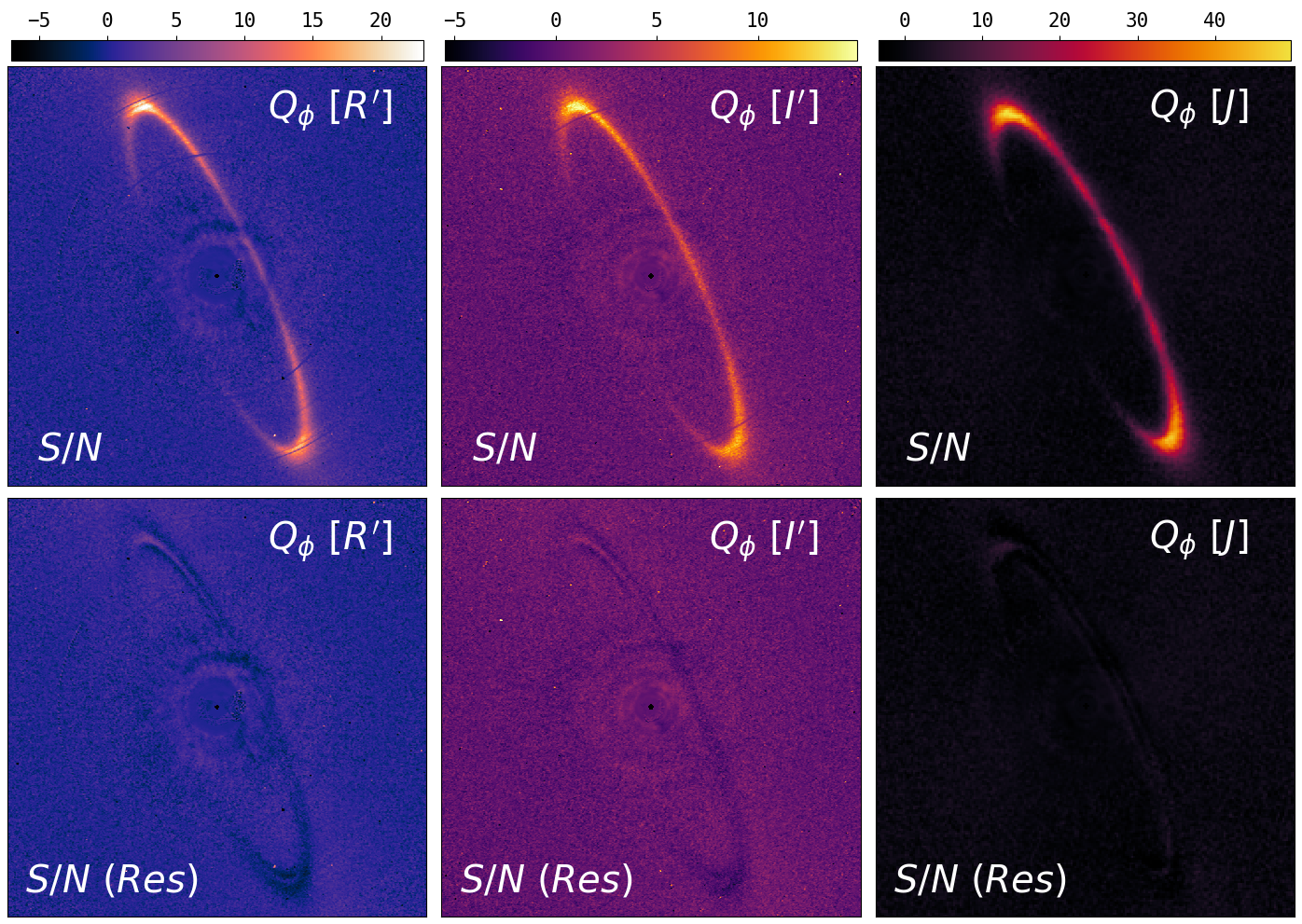}
    \caption{Polarized intensity: S/N map (top) and S/N values of the residuals (bottom).}
    \label{fig_SNR_pol}

\end{figure*}

The top row of fig.\,\ref{fig_SNR_tot} shows the S/N map ($Data/Noise$) of our data (see Fig.\,\ref{fig_Obs_all}), and the bottom row shows the S/N of the residuals ($[Data - BestModel]/Noise$), with the best model shown in Fig.\,\ref{fig_Best_Mod}. Each of the three columns corresponds to the different wavelengths, with, from left to right: ZIMPOL R' band (0.63 µm), ZIMPOL I' band (0.79 µm), IRDIS J band (1.25 µm). For better visibility and to allow a direct comparison, the colormaps are limited between the 0.01\textsuperscript{th} and 99.99\textsuperscript{th} percentile of the total intensity S/N map at each wavelength (both for the S/N and the S/N (res) maps). Fig.\,\ref{fig_SNR_pol} shows the same thing for the polarized data, with the colormaps limited between the 0.01\textsuperscript{th} and 99.99\textsuperscript{th} percentile of the polarized S/N map. 

\end{appendix}
\end{document}